\documentstyle[psfig]{mn2e}

\newif\ifAMStwofonts

\newcommand{\lapp}{\mbox{\raisebox{-0.3em}{$\stackrel{\textstyle <}{\sim}$}}}
\newcommand{\gapp}{\mbox{\raisebox{-0.3em}{$\stackrel{\textstyle >}{\sim}$}}}

\title[Rejuvenated radio galaxies J0041+3224 and J1835+6204]{Rejuvenated radio galaxies J0041+3224 and J1835+6204: how long 
                                                             can the quiescent phase of nuclear activity last?}
\author[C. Konar et al.]
       {C. Konar$^1$ $\thanks{E-mail: chiranjib.konar@gmail.com (CK)}$,  M.J. Hardcastle$^2$,   M. Jamrozy$^3$, J.H. Croston$^4$, S. Nandi$^5$ \\
$^1$Institute of Astronomy and Astrophysics, Academia Sinica, National Taiwan University, Taipei 10617, Taiwan, R.O.C.  \\
$^2$ School of Physics, Astronomy and Mathematics, University of Hertfordshire, College Lane, Hatfield AL10 9AB, UK. \\
$^3$ Obserwatorium Astronomiczne, Uniwersytet Jagiello\'nski, ul. Orla 171, 30244 Krak\'ow, Poland \\
$^4$ School of Physics and Astronomy, University of Southampton, Southampton SO17 1BJ, UK. \\ 
$^5$ Aryabhatta Research Institute of Observational Sciences, Manora Peak, Nainital 263 129, India \\ 
}
\date{Accepted.    Received }

\pagerange{\pageref{firstpage}--\pageref{lastpage}}
\pubyear{  }

\begin{document}

\maketitle

\label{firstpage}

\begin{abstract}
We present radio observations of two well-known double-double radio
galaxies, J0041+3224 and J1835+6204, at frequencies ranging
from 150 to 8460 MHz, using both the Giant Metrewave Radio Telescope
and the Very Large Array. These observations, over a large
radio frequency range, enable us to determine the spectra of the inner
and outer lobes. Our detailed spectral ageing analysis of their inner
and outer lobes demonstrates that the outer doubles of double-double 
radio galaxies are created by the previous cycle of activity, while 
the inner doubles are due to the present cycle of activity. The 
(core subtracted) spectra of the inner doubles of both sources are 
power laws over a large frequency range. We found that the duration 
of the quiescent phase of J0041+3224 is between 4 and 28 per cent of 
the active phase of the previous activity. The outer north-western 
lobe of J1835+6204 has a compact hotspot and the regions of both the 
outer hotspots have close to power-law (rather than curved) spectra, 
which indicates that the outer lobes are still fed by jet material 
ejected in the previous episode just before the central engine stopped 
powering the jet. We estimate that the duration of the quiescent phase 
of J1835+6204 is $\lapp$5 per cent of the duration of the active phase 
of the previous activity. Therefore, we conclude that the duration of 
the quiescent phase can be as short as a few per cent of the active 
phase in radio galaxies of this type.
\end{abstract}


\section{Introduction}
It is established beyond doubt that jet forming activity in radio
galaxies is episodic in nature.  This episodic jet-forming activity
often gives rise to Double-Double Radio Galaxies (DDRGs), which are
defined to be those having a pair of double radio sources with a
common centre, and are thought to occur when a new epoch of jet
activity is triggered in a radio galaxy with older lobes still visible
from the activity of the previous epoch (Schoenmakers et
al. 2000a). At present, about two dozen DDRGs are known in the
literature (see Saikia \& Jamrozy 2009 for a review). Although the
DDRGs are more common, in fact more than two episodes of jet forming
activity are physically plausible: Brocksopp et al. (2007) reported
the first example of three such episodes in a radio galaxy, and
recently Hota et al. (2011) reported a possible case of a
triple-double radio galaxy dubbed `SPECA' (J1409$-$0302).

The very existence of well-shaped and well confined radio galaxy lobes
suggests that there must be some medium confining the lobes either
through ram pressure or thermal pressure. Even radio galaxies whose 
optical environments are known to be comparatively poor are found to 
possess X-ray-emitting hot gas environments, corresponding to group or
cluster-scale X-ray luminosities, whose gas properties can now be 
modelled in detail (see Hardcastle \& Worrall, 1999; Hardcastle \& Worrall, 
2000; Worrall \& Birkinshaw, 2000; Croston et al., 2003; Croston et al., 
2004; Evans et al., 2005; Belsole et al., 2007; Croston et al., 2008; 
Konar et al., 2009)
This environment emits via thermal bremsstrahlung in the
X-ray band, and can be classified as a poor-cluster to group scale
environment. The magnetised relativistic plasma (MRP) of the radio
lobes does not mix well with the external thermal gas in the
environment, as can be inferred from limits on internal depolarization
of lobes at low frequencies, nor does it diffuse out completely into
the IntraCluster Medium (ICM) even long after the jet stops feeding
the lobes. Kaiser, Schoenmakers \& R\"{o}ttgering (2000) argued
against the entrainment of material from the surrounding hot-gas
environment through the lobe periphery. However, the observed
well-confined inner lobes of DDRGs are not capable of existing in such
a tenuous cocoon of matter deposited in the previous cycle of jet
activity (the typical number density of relativistic radiating
particles is $\sim10^{-10}$ cm$^{-3}$). Therefore, Kaiser et
al.\ (2000) proposed that the ambient thermal medium is a two phase
medium. One phase consists of dense warm clouds and the other phase is
the hot-gas medium, which has the characteristics of the general ICM,
with a temperature of $\sim 10^7$K and a central particle number
density of $10^{-1}-10^{-3}$ cm$^{-3}$. The dense, warm cloud phase
has a temperature of $\sim 10^4$K and a particle number density of
$\sim$100 cm$^{-3}$, with a very low filling factor of
$10^{-8}-10^{-3}$, dependent on distance from the Active Galactic 
Nucleus (AGN). The hot-gas medium has a filling factor $\sim 1$.  
Kaiser et al. (2000) discussed in detail, in Section~2 of their paper, 
the observational evidence for these warm clouds through optical emission 
line detections. The warm clouds are dynamically unimportant, as their
filling factor is extremely small. When the radio lobes expand very
rapidly, these clouds are not efficiently accelerated by the lobe
expansion, being heavier than the hot-gas phase, and so are overtaken
by the contact discontinuity. This means that the warm clouds become
dispersed inside the lobe and provide the required matter density to
prevent the inner jets from being ballistic. In this model, to
understand the dynamics of these sources it is necessary to know the
pressure and matter density internal and external to the lobes.

The condition for a radio galaxy to restart is not well
understood. Schoenmakers et al.\ (2000a) explored the possibility of
causing restarting jets in DDRGs by interaction and merger events,
although there is no observational evidence for this model. Numerical
simulations of colliding galaxies show that these usually do not merge
completely in the first encounter (Barnes \& Hernquist,
1996). According to this scenario, the in-falling galaxy loses a large
fraction of its gas and stars to the main galaxy, while parts of it
pass through the main galaxy. These parts will turn around and collide
again with the main galaxy. If the first encounter triggers the AGN,
then the subsequent encounters destabilise it and jet interruption
occurs. However, a problem with this scenario is that, as argued by
Schoenmakers et al. (2000a), it would lead to the expectation of
finding many more restarting radio galaxies than we see today.  The
mechanism for triggering multiple episodes of jet activity could also
be closely related to the feedback mechanism proposed for solving the
cooling flow problem in clusters of galaxies (see McNamara \& Nulsen,
2007).  Recently, the study of Jetha et al. (2008) showed that the
restarting phenomenon in B2 0838+32A is consistent with
feedback-driven accretion by the AGN. Therefore, studying DDRGs is
very important to understand the conditions for the phenomenon of
restarting jets, their relation to feedback mechanisms, and the
dynamics of radio galaxies in general.  

  To understand the dynamics of radio galaxies we need to know the
  injection spectral index (which is related to the particle
  acceleration process), the spectral age (which is a good
  approximation to dynamical age), the jet power (which is connected
  to the accretion rate at the central engine), the magnetic field
  strength based on the equipartition assumption (which we know is
  reasonably accurate in FR\,II radio galaxies: Croston et al.\ 2005), 
  and the internal pressure of the lobes (which helps us determine the 
  composition of the relativistic particles, by comparing with the 
  external pressure obtained from X-ray observations); all these 
  parameters can be estimated from radio observations.  Therefore, 
  comprehensive radio studies of such sources are essential to pave 
  the way towards a complete understanding of the dynamics of radio 
  galaxies. {\bf Our further work related to the time-scale of quiescent 
  phase of episodic radio galaxies, injection index in different 
  episodes and particle acceleration at hotspots, and the dynamics of 
  DDRGs through radio and X-ray observations will be published soon
  by Konar et al. (in prep); Konar \& Hardcastle (in prep) 
  and Konar et al. (in prep) respectively.}

In this paper, we have carried out a multifrequency radio study of two
DDRGs, namely J0041+3224 and J1835+6204.  We present the Giant Meterwave 
Radio Telescope (GMRT) and Very Large Array (VLA) observations in Section~2, 
our observational results in Section~3, our spectral ageing analysis in 
Section~4, a discussion in Section~5 and concluding remarks in Section~6.

The cosmological parameters that we have used are $H_o=71$ km s$^{-1}$
Mpc$^{-1}$, $\Omega_M=0.27$ and $\Omega_{vac}=0.73$ (Spergel et al.
2003). In this cosmology, 1 arcsec corresponds to 5.733 kpc for the
source J0041+3224 situated at an (estimated, see below) luminosity
distance $D_L=2486.3$ Mpc. For J1835+6204, 1 arcsec corresponds to
6.207 kpc based on its luminosity distance of $D_L=2955.7$ Mpc.

\section{Observations and data reduction}
The images published in this paper are from the GMRT and the VLA. 
All GMRT data are from our observations with project code 10CKa01. 
{\bf All VLA data are public data from the VLA archive.} The details of 
the observations with both telescopes are given in Table~\ref{obslog}. 

The shortest baseline for the C and D configuations of the VLA is
35 m. Our target sources, J0041+3224 and J1835+6204 have sizes
$\sim 2.8$ arcmin and $\sim 3.7$ arcmin respectively; however, in both
sources, at high frequencies, no single structure with dominant flux
density is larger than 2 arcmin. As all the VLA maps (published by
Saikia et al. 2006) used for flux density measurements of J0041+3224
were made from C-array data, where the largest angular size that can
be mapped without loss of flux is 3 arcmin, our measurements for
J0041+3224 should not be susceptible to loss of flux due to lack of
short spacings in the uv-coverage.  Alhough the 8.4 GHz map of
J1835+6204 (Figure~\ref{image.j1835}) looks continuous from hotspot to
hotspot, this is (at least partially) due to the effects of
comparatively poor resolution, which is why we do not see such
continuity in the 4.8 GHz images at somewhat higher resolution. The
8.4-GHz image of J1835+6204 is made by us from a combination of
observations made separately with the B and D configurations, which
again can image 3 arcmin regions without loss of flux. So this source
is also not susceptible to any major loss of flux due to the lack of
short spacings in the uv coverage. The shortest baseline of the GMRT
is 100 m, and so we can safely map emission regions of a size of 7 arcmin
at 1280 MHz without loss of flux; therefore, the GMRT can map
even larger structure at any frequencies below 1280 MHz without losing
any flux. We conclude that neither the VLA nor the GMRT images suffer from
any significant loss of flux due to the lack of short spacings. In
addition, we see no observational signatures of such flux losses, and
so conclude that our flux measurements and spectral analysis is
reliable.

\subsection{GMRT observations}
The observations were made in the standard manner, with each
observation of the target source interspersed with observations of
calibrators. For an observing run of any target source, one of 3C48,
3C147 or 3C286 was observed as a flux density and bandpass calibrator
at a given frequency. At each frequency the source was observed in a
full-synthesis run of approximately 9 hours including calibration
overheads. On-source durations vary from 240$-$371 min. Full details
of the GMRT array can be found at {\tt
  http://www.gmrt.ncra.tifr.res.in}.  The data collected were
calibrated and reduced in the standard way using the NRAO {\tt AIPS}
software package. Flux density calibration of our GMRT data uses the scale
of Baars et al. (1977). All of the GMRT data below L band were
affected by Radio Frequency Interference (RFI) to various extents, and
were extensively flagged by hand within {\tt AIPS} to reduce the
effects of RFI as much as possible.

\subsection{VLA observations}
The VLA data for the sources at various frequencies ($\sim$1.4,
4.8, 8.4 GHz) were analysed to produce images. The data from our
own observations and those from the archive are indicated in the
observation log table (Table~\ref{obslog}).  All VLA observations were
in snap-shot mode. The on-source observing time was between 7 to 251
min (see Table~\ref{obslog}). One of the calibrators 3C48, 3C147 or 3C286 was observed
as a flux density calibrator at each frequency of a given observing
run. As in the case of the GMRT data, the VLA data were edited and
reduced using the {\tt AIPS} package. The use of VLA archival data to
supplement our GMRT observations was essential to constrain the
spectra of the DDRGs and to allow us to carry out the spectral ageing
analysis. All flux densities are on the Baars et al. (1977) scale.

\begin{figure}
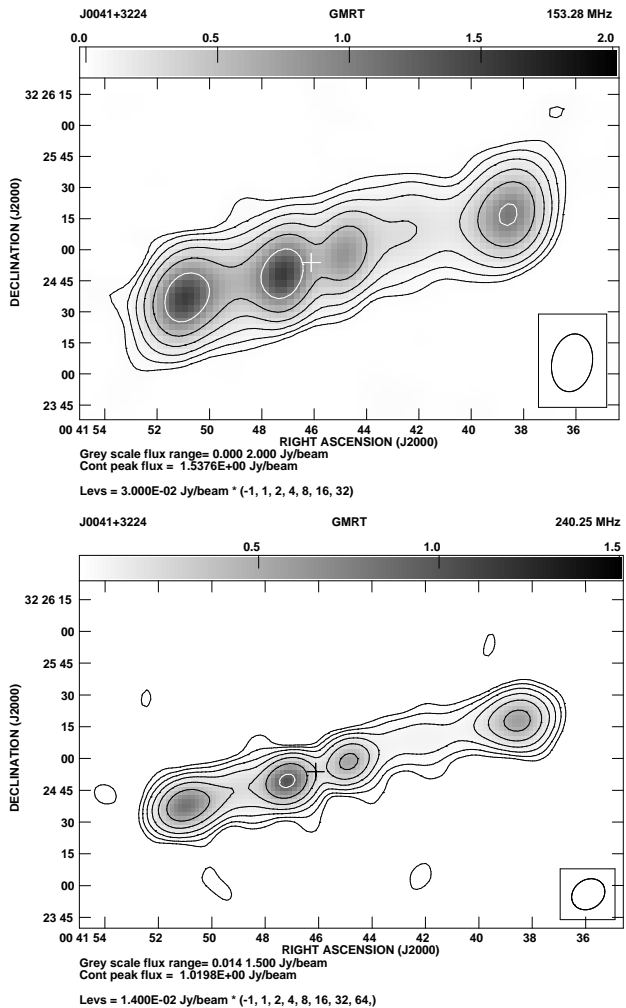

\vbox{
   \psfig{file=J0041+3224_E_W.GREY.SCALE.PS,height=2.65in,angle=-90}
   \psfig{file=J0041+3224_T_W.GREY.SCALE.PS,height=2.65in,angle=-90}
}
\caption[]{Full resolution radio images of J0041+3224 are shown. The frequency of each image and the
telescope with which the image is made are given at the top of each image. The peak flux density, greyscale level, 
first contour and the contour levels are all given at the bottom of each image. A $+$ sign indicates the position 
of the optical galaxy. We have not shown the 325 MHz image as its quality is not good enough. However, the flux density 
seems to be consistent when compared with all other frequencies. The rms noises and beam sizes are given in 
Table~\ref{obsparam.j0041}. 
  }
\label{image.j0041}
\end{figure}
\section{Observational results}
\subsection{Observed parameters and structures}
\label{structure}
The images of the entire source using the GMRT and the VLA are
presented in Figures \ref{image.j0041} and \ref{image.j1835}, while
the observational parameters and some of the observed properties are
presented in Table \ref{obsparam.j0041} and \ref{obsparam.j1835}. The
flux densities of different components were  estimated by
specifying an area around each component. {\bf All the flux density 
measurements have been done with the {\tt AIPS} task `TVSTAT' which 
requires selection of a shaded polygonal area in the region of interest.
`TVSTAT' adds up all the pixel values within the shaded area to give 
the total flux density of the shaded area.} We measured the flux density
of each component using several different areas around the radio
emitting component, which gives an idea of the systematic errors due
to our choice of region. The change in flux density of each component,
when measured this way, is only a few per cent.  Flux density
  values that have been measured directly by us from the FITS maps
  with {\bf the {\tt AIPS} task `TVSTAT'} are assumed to have the 
  following error values: 5 per cent
  for 1400, 4860 and 8460 MHz VLA measurements; 7 per cent for 1287
  and 610 MHz GMRT measurements; and 15 per cent for 332, 240 and 153
  MHz GMRT measurements. Other flux density values quoted in the paper
  have been assigned errors determined by propagating errors from the
  errors of directly measured flux densities.
\begin{figure*}
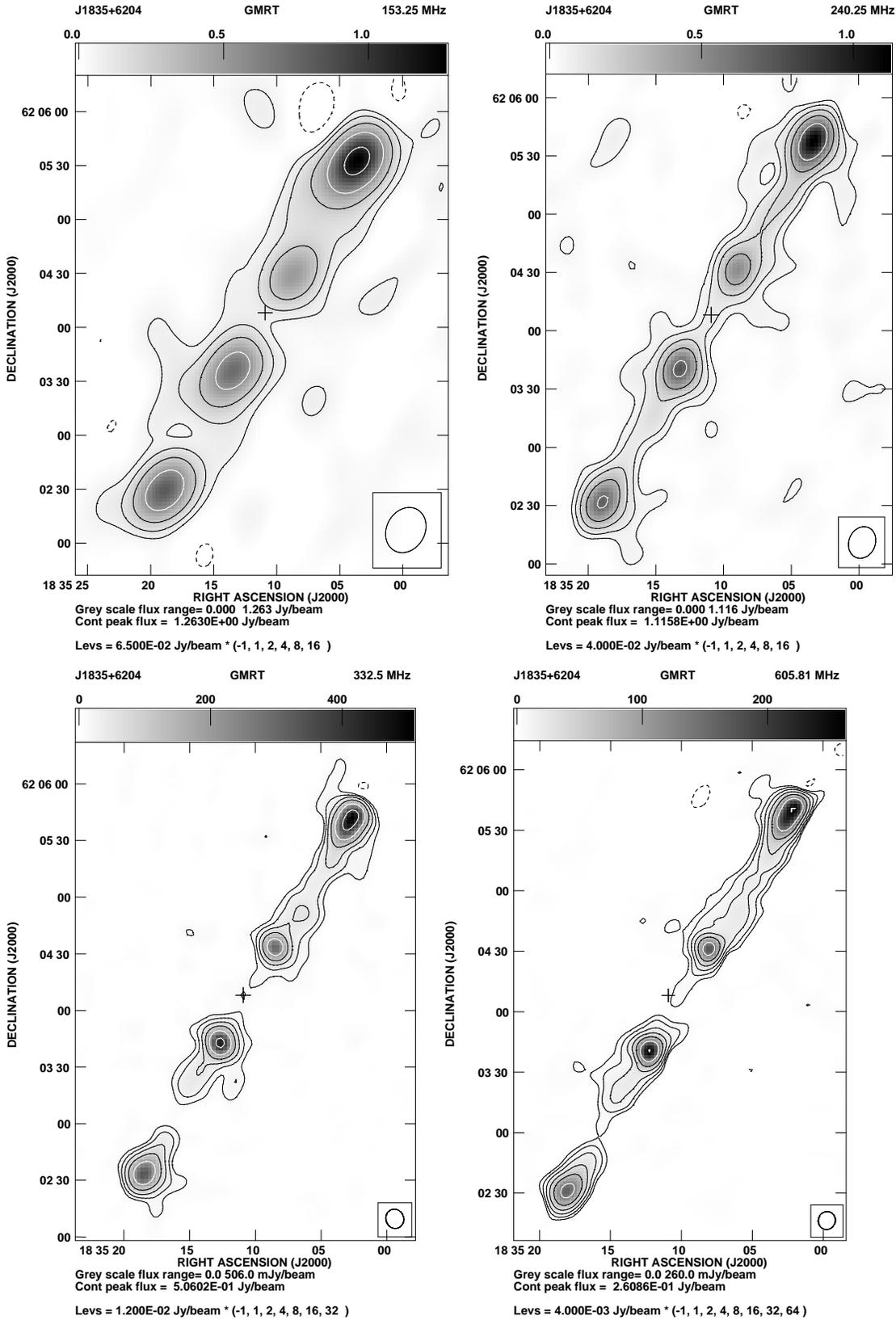

\vbox{
\hbox{
 \psfig{file=J1835+6204_E.PS,height=4.3in,angle=0}
 \psfig{file=J1835+6204_T.PS,height=4.3in,angle=0}
} 
\hbox{
 \psfig{file=J1835+6204_P.PS,height=4.3in,angle=0}
 \psfig{file=J1835+6204_G_SN.PS,height=4.3in,angle=0}
}
} 
\caption[]{Full resolution radio images of J1835+6204 are shown. The exact frequency, and the name of the 
telescope (and observation codes of the VLA data sets used to produce the images) are
given at the top of each image.  Some of the VLA images are produced by combining two data sets to detect the diffuse emission.  
The peak flux density, greyscale level, 1st contour and the contour levels are all given at the bottom of each image. A $+$ sign
indicates the position of the optical galaxy. The rms noises and the beam sizes are given in
Table~\ref{obsparam.j1835}.
}
\label{image.j1835}
\end{figure*}
\begin{figure*}
\vbox{
\hbox{
   \psfig{file=J1835+6204_GMRT.L.PS,height=4.5in,angle=0}
   \psfig{file=J1835+6204_AL412A.L.PS,height=4.5in,angle=0}
}
\hbox{
   \psfig{file=J1835_AL412A.AM954_C.PS,height=4.5in,angle=0}
   \psfig{file=J1835+6204_AS586B.X.PS,height=4.5in,angle=0}
}
}
\contcaption{ }
\end{figure*}

\begin{figure*}
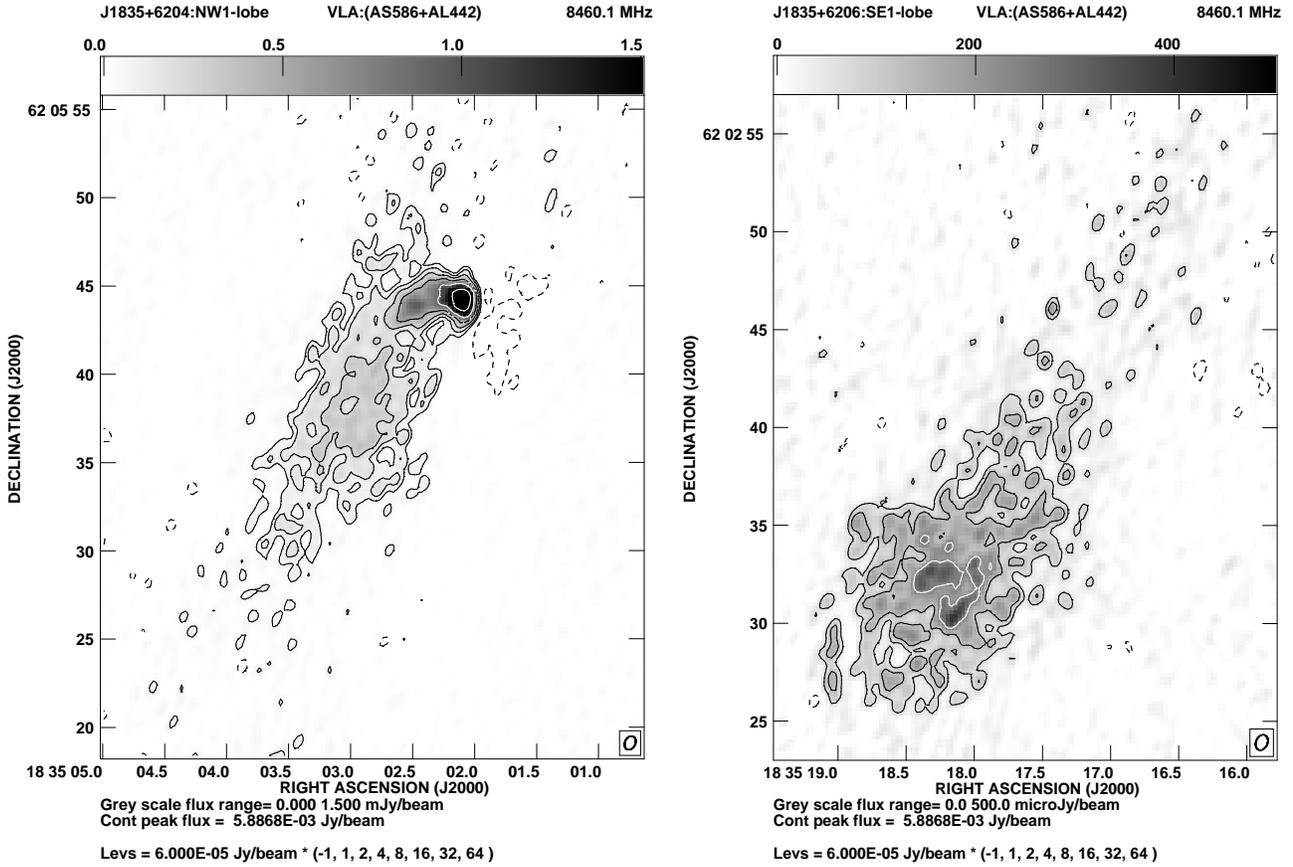

\vbox{
\hbox{
   \psfig{file=J1835+6204_AS586BAL442A.NW1.LOBE.X.PS,height=4.6in,angle=0}
   \psfig{file=J1835+6204_AS586BAL442A.SE1.LOBE.X.PS,height=4.6in,angle=0}
}
}
\caption[]{High resolution X band radio images of NW1 and SE1 lobes of J1835+6204 are shown. The images of two lobes of this 
source is from the same map produced from the data obtained by combining two separate data sets.  
The resolution of these images is $0.82\times0.59$ arcsec$^2$ at $PA\sim336^{\circ}$ and the rms noise is 21 $\mu$Jy/beam. 
The exact frequency, and the name of the telescope and the observation codes of the data sets used to produce this map are 
given at the top of each image. The peak flux density, greyscale level, 1st contour and the contour levels are all given 
at the bottom of each image.
}
\label{image.j1835_X-band}
\end{figure*}
\begin{table*}
\caption{Observing log:
Column 1 shows the source name; columns 2 and 3 show the name of the telescope, and the array configuration for 
the VLA observations; column 4 shows the frequency of the observations; while bandwidth and on-source observing time
have been listed in column 5 and 6, respectively. The dates of the observations are listed in 
column 7, and column 8 shows the project code of the data. All GMRT data are our observations. 
}
\label{obslog}
\begin{tabular}{l c c l l r  l c}
\hline
Source      &   Teles-    & Array  & Obs.      & Bandwidth & On-source &  Obs. Date  &  Project  \\
            &   cope      & Conf.  & Freq.     & used for  & observing &             &  code     \\
            &             &        &           &  mapping  & time      &             &           \\      
            &             &        & (MHz)     & (MHz)     &  (min)    &             &           \\
 (1)        &   (2)       &  (3)   & (4)       &  (5)      &   (6)     &   (7)       &  (8)      \\
\hline                                                                       
J0041+3224  &   GMRT      &        & 153.28    &   5.00    &  360      &11-AUG-2006  &  10CKa01  \\
            &   GMRT      &        & 240.25    &   5.00    &  307      &17-JUL-2006  &  10CKa01  \\    
            &   GMRT      &        & 331.88    &  12.50    &  270      &15-JUN-2006  &  10CKa01  \\  

J1835+6204   & GMRT        &       & 153.25     &   5.00    &240        &01-JUN-2006  & 10CKa01   \\
             & GMRT        &       & 240.25     &   5.00    &280        &14-MAY-2006  & 10CKa01   \\
             & GMRT        &       & 332.50     &  12.50    &371        &16-JUN-2006  & 10CKa01   \\
             & GMRT        &       & 605.81     &  12.50    &280        &14-MAY-2006  & 10CKa01   \\
             & GMRT        &       &1287.50     &  12.50    &360        &13-JUL-2006  & 10CKa01   \\
             & VLA         &   B   &1442.50     &  50.00    &  8.0      &18-FEB-1997  & AL412A    \\ 
             & VLA         &   A   &1440.50     &  50.00    &  251      &21-OCT-2000  & AS697     \\ 
             & VLA         &   D   &4860.10     & 100.00    & 14        &27-JUN-2008  & AM954     \\ 
             & VLA         &   B   &4860.10     & 100.00    & 7.0       &18-FEB-1997  & AL412A    \\ 
             & VLA         &   D   &8460.10     & 100.00    & 23        &23-AUG-1996  & AS586     \\ 
             & VLA         &   B   &8460.10     & 100.00    & 30        &07-OCT-1998  & AL442     \\ 
\hline
\end{tabular}
\\
\end{table*}

For both the sources the component designations are as follows. The
outer north-western and south-eastern lobes are called NW1 and SE1,
respectively. The inner north-western and south-eastern lobes are
called NW2 and SE2, respectively. Since in low frequency images the
inner doubles appear embedded in the diffuse plasma of the outer
cocoons, we could not measure the flux densities of the inner and
outer lobes separately on a particular side of the core. In these
cases, we measured the total flux densities of the entire NW1 plus NW2
lobes and SE1 plus SE2 lobes, and designated those joint components as
the NW1\&2 and SE1\&2 components respectively.  All the inner lobes
have edge-brightened structures called hotspots, and hence are of FRII
type in morphology.  The outer lobes of J0041+3224, as seen in high
resolution images, do not have any compact hotspots, but neither do
their structures resemble an FRI morphology. In the poor resolution
images we see a peak of emission in the outer lobes at both ends of
the source along the length. These are most probably the locations of
the hotspots when the outer sources were active. These locations can
be called warm-spots. Therefore, the outer structures are consistent
with the idea that they are created in the previous episode of Jet
Forming Activity (JFA). 

The case of J1835+6204 is different. At the outer end of the NW1 lobe
of J1835+6204, we see a reasonably compact component (hotspot), though
the SE1 lobe does not show such a component. However, the radio
spectra at the position of the outer hotspots are well described by
power laws (see Figure~\ref{strip.fit_j1835}). This means that the outer
lobes at the location of their hotspots are still fed by jet material
(for NW1 lobe) or else the time elapsed since the last jet material
reached the outer hotspot is short (for the SE1 lobe). In other words,
the spectra have not yet suffered sufficient losses to develop
curvature. Both inner and outer pairs of hotspots are more or less 
co-linear with the core, and both the doubles are also reasonably 
aligned with each other.

\subsection{Notes on the sources}
J0041+3224 (B2\,0039+32): this DDRG has a very elongated and
cylindrical shape. The projected linear sizes of the inner and outer
doubles are 171 and 969 kpc respectively. This makes the outer
structure a giant radio galaxy. The host galaxy of this source is a
20th magnitude galaxy [NASA/IPAC Extragalactic Database (NED)]. This source does
not have a measured redshift; however, we adopt the estimated redshift
of 0.45 quoted by Saikia et al. (2006). In low frequency images we can
see that the inner double is ploughing its way through the cocoon
material of the outer lobes. The inner lobes have compact hotspots
{\bf (see Saikia et al. 2006)} which are characteristic of FRII radio 
galaxies. The outer doubles do not have any hotspots. Radio images at two 
frequencies are shown in Figure~\ref{image.j0041} \\

\noindent
J1835+6204 (B1834+620): this DDRG also has a very elongated and 
cylindrical shape. The projected linear sizes of the inner and outer 
doubles are 369 and 1379 kpc, respectively. The outer structure of this 
radio galaxy is also of giant size. Its host galaxy is a 19.7th magnitude
galaxy with a redshift of $0.5194\pm0.0002$ (Schoenmakers et al. 2000a). 
The inner double of this DDRG also appears to be ploughing
through the cocoon plasma of previous JFA. Unlike J0041+3224, one of
the outer lobes (NW1 lobe) of this sources has a compact hotspot-like
feature. From Figure~\ref{image.j1835_X-band}, it is clear that the jet 
of NW1 lobe in the previous episode had bent in the western direction 
by 90 degree with respect to the original jet axis. Figure~\ref{image.j1835} 
and \ref{image.j1835_X-band} display the images of this source at 
various radio frequencies and resolutions.

\begin{table*}
\scriptsize{
\caption{The observational parameters and flux densities of the outer and inner lobes of J0041+3224. 
The description of the columns are as follows. Column 1 shows frequency of observation; columns 2, 3 and 4 show the beam 
size and its orientation; column 5 lists the rms noise of the map; {\bf column 6 presents the integrated flux density}; 
columns 7, 10, 13, 16 list the component designation. Columns 8, 11, 14 and 17 list the peak flux density of the component. 
Columns 9, 12, 15 and 18  list the flux density of the component. The component designations are explained in Section\ref{structure}.
   }
\label{obsparam.j0041}
\begin{tabular}{l rrr r r l rr l rr l rr l rr} 
\hline
Freq.        & \multicolumn{3}{c}{Beam size}& rms     & S$_I$   & Cp      & S$_p$ & S$_t$  & Cp    & S$_p$ & S$_t$ & Cp       & S$_p$   & S$_t$   &  Cp & S$_p$ & S$_t$       \\
MHz          & $^{''}$ & $^{''}$& $^\circ$ &    mJy   & mJy     &         & mJy   & mJy    &       & mJy   & mJy   &          & mJy     & mJy     &     & mJy   &             \\
             &         &        &          &     /b   &         &         & /b    &        &       & /b    &       &          & /b      &         &     &  /b   &             \\
(1)          & (2)     & (3)    & (4)      &    (5)   & (6)     & (7)     & (8)   & (9)    & (10)  & (11)  & (12)  & (13)     & (14)    & (15)    & (16)& (17)  & (18 )       \\
\hline \hline

G153.28      & 28.14   & 19.55  &  349     &   8.88   &  7095   &SE1\&2    & 1538  & 4250   &       &       &       & NW1\&2  &  1054   & 2927    &     &       &             \\
            
G240.25      & 16.58   & 13.66  &  304     &   4.11   &  4374   &SE1\&2    & 1020  & 2615   &       &       &       & NW1\&2  &   581   & 1752    &     &       &             \\
                    
G331.88      & 09.46   & 08.41  &   61     &   4.87   &  4206   &SE1\&2    &  960  & 2465   &       &       &       & NW1\&2  &   493   & 1640    &     &       &             \\
G616.81      &  6.50   & 4.80   &  165     &   0.49   &  2211   &SE1\&2   &  567  &  1289  &       &       &       & NW1\&2   &  274    &  856    &      &      &            \\

G1287.44     &  2.61   & 2.33   &   25     &   0.20   &  1016   &SE1      &  13   &  232   & SE2   &  231  & 373   & NW1      &   12    &  262    & NW2  & 92   & 145        \\   

V1400.00     & 14.65   & 12.64  &   11     &   0.13   &   940   &SE1\&2   &  338  &  582   &       &       &       & NW1\&2  &   139   &  360    &      &      &            \\
                                                                                                                             
V4860.10     &  4.04   & 3.65   &   13     &   0.03   &   296   &SE1      &  6    &  36    & SE2   &  126  & 154   & NW1      &   7.4   &   49    & NW2  & 42   &  52        \\
                                                                                                                             
V8460.10$^{\dag}$ &  2.41 & 2.26 &   19    &   0.02   &   160   &SE1      &   1   &  9.31  & SE2   &  75   & 100   & NW1      &   1.8   &   16    & NW2  & 21   &  32        \\
\hline
\end{tabular}
\\
\begin{flushleft}
$^{\dag}$: the flux densities are measured after primary beam correction, which was not done in Saikia et al. 2006. \\
\end{flushleft}
}
\end{table*}

\begin{table*}
\scriptsize{
\caption{ The observational parameters and flux densities of the outer and inner lobes of J1835+6204. 
The column descriptions and the component designations are exactly the same as in Table~\ref{obsparam.j0041}.
}
\label{obsparam.j1835}
\begin{tabular}{l rrr r r l rr l rr l rr l rr}
\hline
Freq.       & \multicolumn{3}{c}{Beam size}                    & rms      & S$_I$   & Cp  & S$_p$  & S$_t$  & Cp   & S$_p$ & S$_t$ & Cp  & S$_p$   & S$_t$   &  Cp  & S$_p$ & S$_t$   \\

            MHz         & $^{\prime\prime}$ & $^{\prime\prime}$ & $^\circ$ &    mJy   & mJy     &     & mJy    & mJy    &      & mJy   & mJy   &     & mJy     & mJy   &   & mJy & mJy    \\
                        &                   &                   &          &  /b      &         &     & /b     &        &      & /b    &       &     & /b      &        &   &  /b &       \\
(1) & (2) & (3) & (4) & (5) & (6) & (7) & (8) & (9) & (10) & (11) & (12) & (13) & (14) & (15) & (16) & (17) & (18 ) \\
\hline \hline

 G153.25      & 26.18   & 21.27  &  331      &   13.1   &  5510   &SE1\&2  &  811  &  2539  &       &       &       & NW1\&2  &   1263  &  2898    &     &     &         \\

 G240.25      & 16.68   & 13.71  &  337      &   7.31   &5762$\dag$&SE1\&2  &  740  &  2653  &       &       &       & NW1\&2  &   1116  &  3141    &     &     &         \\

 G332.50      & 10.28   & 09.51  &  21.5     &   1.72   &  3375   &SE1\&2  &  441  &  1555  &       &       &       & NW1\&2  &    506  & 1813     &     &     &         \\

 G605.81      &  8.94   & 8.40   &  341      &   1.25   &  1813   &SE1\&2  &  261  &   800  &       &       &       & NW1\&2  &    257  &  980     &     &     &         \\
 
 G1287.50     &  2.99   & 2.40   &  338      &   0.17   &   862   &SE1\&2  &  102  &   390  &       &       &       & NW1\&2  &     70  &  462     &     &     &         \\
 V1442.50$^a$ &  4.99   & 3.66   &  334      &   0.02   &   708   &SE1\&2   & 101   &   318  &       &       &       & NW1\&2  &   75    &  388     &     &     &         \\

 V1440.50$^b$ &  1.35   & 1.00   &  357      &   0.04   &   657   &SE1      & 2.7   &   153  & SE2   &  47   &  135  & NW1      &   23    &  273     & NW2 &  30 &  83     \\

 V4860.10$^a$ &  1.52   & 1.12   &  324      &   0.02   &   200   &SE1      & 1.1   &    37  & SE2   &  17   &   46  & NW1      &   11    &   76     & NW2 &  13 &  33     \\
 
 V4860.10$^c$ & 15.31   & 10.03  &   46      &   0.03   &   245   &SE1\&2   & 44    &   110  &       &       &       & NW1\&2  &   48    &  133     &     &     &         \\
  
 V4860.10$^d$ &  1.71   &  1.14  &  322      &   0.02   &   257   &SE1      & 2.1   &    70  & SE2   &  21   &   50  & NW1      &   12    &   96     & NW2 &  15 &  32     \\

 V8460.10$^e$ &  0.79   &  0.57  &  336      &   0.02   &    75   &SE1      & 0.20  &   5.6  & SE2   &  5.5  &   26  & NW1      &   2.9   &   26     & NW2 & 3.4 &  18     \\

 V8460.10$^f$ &  8.79   &  6.45  &  14.5     &   0.04   &   148   &SE1\&2   & 26    &   68   &       &       &       & NW1\&2  &   22    &   78     &     &     &         \\   

 V8460.10$^g$ &  0.82   &  0.59  &  336      &   0.02   &   141   &SE1      & 0.4   &   34   & SE2   &  5.9  &   30  & NW1      &   3.1   &   53     & NW2 & 3.7 &  18     \\  

\hline
\end{tabular}
\\
\begin{flushleft}
$\dag$: Comparing with the flux density values at all other frequencies, it is likely to have some systemic error.
$^a$: Project code- AL412A, $^b$: Project code- AS697, $^c$: Project code- AM954, $^d$: Project code- (AL412A+AM954), map from combined data,
$^e$: Project code- AL442,  $^f$: Project code- AS586, $^g$: Project code- (AS586+AL442A), map from combined data.  \\

\end{flushleft}
}
\end{table*}
\begin{table*}
\scriptsize{
\caption{The flux densities of the inner doubles of J0041+3224 and J1835+6204 from our measurements.
The common lower uv cut-off of 3.0 k$\lambda$ has been applied to the data sets to image the inner
double free from outer diffuse emission. The GMRT L-band and VLA high-frequency data have
been used to image the inner doubles to constrain the spectra of the components of the inner
doubles. The description of the table is as follows: column 1: frequency of observations
with the name of the source; columns 2-4: major axis, minor axis and position angle of the
synthesized beam; columns 5, 7 and 9: total flux density of inner double (without core),
flux density of inner SE lobe and that of inner NW lobe; columns 6, 8 and 10: the errors of
the flux densities; column 11: project code of the data used.}
\label{tab_inn.dbl.flux}
\begin{tabular}{l l l l l l r c r r c r r r}
\hline
Freq.          &  \multicolumn{3}{c}{Beam size}  & $S^{tot}_{inn}-c$ &  Err    & $S_{SE2}$ & Err       & $S_{NW2}$   & Err         & Project   \\
(MHz)          & $^{''}$ & $^{''}$ & $^{\circ}$  &  (mJy)            & (mJy)   &  (mJy)    & (mJy)     & (mJy)       & (mJy)       & code      \\
               &         &         &             &                   &         &           &           &             &             &           \\
(1)            &  (2)    & (3)     & (4)         &  (5)              &  (6)    &  (7)      &  (8)      & (9)         &  (10)       & (11)      \\
\hline
J0041+3224    &          &         &             &                   &         &           &           &             &             &           \\  %
1287.5        &  2.61    &  2.33   &    25       & 518               &  28     &   373     &  26       & 145         & 10          &   a       \\  
4860.1        &  4.04    &  3.65   &    13       & 206               &   8     &   154     &   8       &  52         &  3          &   b       \\  
8460.1        &  2.41    &  2.26   &    19       & 132               &   5     &   100     &   5       &  32         &  2          &   b       \\  
J1835+6204    &          &         &             &                   &         &           &           &             &             &           \\  %
1287.5        &  2.10    &  1.84   &  338        & 250               & 12.70   &  154      & 10.78     &  96         & 6.72        &    g      \\  
1440.5        &  1.35    &  1.00   &  357        & 218               &  7.94   &  136      &  6.80     &  82         & 4.10        &    h      \\  
4860.1        &  1.67    &  1.09   &  322        &  79               &  2.84   &   47      &  2.35     &  32         & 1.60        &    i      \\  
8460.1        &  0.82    &  0.59   &  336        &  48               &  1.73   &   29      &  1.45     &  19         & 0.95        &    j      \\  
\hline
\end{tabular}
\begin{flushleft}
a: Target of opportunity observation, b: AS741,
g: 10CKa01, h: AS697, i: AL412A+AM954, j: AS586+AL442
\end{flushleft}
}
\end{table*}


\begin{table*}
\caption{The integrated flux densities, the total flux densities of
  the inner and outer doubles, and the uncertainties on these fluxes
  for J0041+3224. The total flux densities of the outer doubles have
  been estimated by subtracting the flux densities of the inner
  doubles from the integrated flux densities. The errors on the total
  flux densities of the outer doubles are propagated from the errors
  of the inner double and integrated flux densities.  The column
  designations are self evident.}
\label{tab_flux_intinnout.j0041}
\begin{tabular}{l l r r c r r l r r c}
\hline
Freq.    &Cp       & $S_{t}$ &  Error & Cp        & $S_{t}$ & Error & Cp        & $S_{t}$ & Error & Reference \\
 MHz     &         &  mJy    &  mJy   &           & mJy     & mJy   &           & mJy     & mJy   &   and      \\
         &         &         &        &           &         &       &           &         &       & comment    \\    
(1)      &  (2)    &  (3)    & (4)    &  (5)      &  (6)    & (7)   & (8)       & (9)     & (10)  & (11)       \\
\hline
   74.00 & Int     &  10882  & 2176   &           &         &       & E1+W1     & 6727    & 2464  & VLSS     \\
  153.28 & Int     &   7095  & 1064   &           &         &       & E1+W1     & 4402    & 1279  &   1      \\
  240.25 & Int     &   4374  &  656   &           &         &       & E1+W1     & 2603    &  840  &   1      \\
  331.88 & Int     &   4206  &  631   &           &         &       & E1+W1     & 2804    &  760  &   1      \\
  408    & Int     &   2200  &  200   &           &         &       &           &         &       &   2      \\
  616.81 & Int     &   2184  &  153   &           &         &       & E1+W1     & 1289    &  319  &   1      \\
 1287.44 & Int     &   1016  &   71   & E2+W2 &518$^{\ast}$ & 28$^a$& E1+W1     &  498    &   76  &   1      \\
 1400.00 & Int     &    940  &   47   &           &         &       &           &         &       &   3      \\
 1400.00 & Int     &    900  &  100   &           &         &       &           &         &       &   4      \\
 1400.00 & Int     &    967  &   48   &           &         &       &           &         &       &   NVSS   \\
 1400.00 &Int (avg)&    936  &   40   &           &         &       & E1+W1     &  441    &  167  &   5      \\   
 4830$^p$& Int     &    291  &   40   &           &         &       & E1+W1     &   89    &   82  &   6       \\
 4850     & Int     &    269  &   15   &           &         &       &            &         &       &   7        \\
 4850     & Int     &    273  &   35   &           &         &       &            &         &       &   8        \\
 4850$^q$ &Int (avg)&    271  &   19   &           &         &       & E1+W1      &   70    &   73  &   9        \\
4840 (Avg)&         &         &        &           &         &       &E1+W1 (Avg) &   80    &   27  &  10        \\
 4860.10  & Int     &    296  &   15   & E2+W2 &206$^{\ast}$ & 08$^b$ & E1+W1     &   90    &   17  &   1      \\
 8460.10  & Int     &    160  &   08   & E2+W2 &132$^{\ast}$ & 05$^b$ & E1+W1     &   28    &   9.4 &   1      \\
\hline
\end{tabular}
\begin{flushleft}
(avg): The average flux density of the values in the previous rows 
  for the same frequency. \\ (Avg): The average of $^p$ and $^q$ \\ 
  $^{\ast}$: The power law spectra of the total inner lobes 
  (without the core) were constrained by a least-squares fit with 
  these data, and extrapolated to the lowest frequency. The flux 
  density of total inner double (core subtracted) at all other 
  frequencies was estimated from the fitted power law. \\
$^a$: Fluxes of E2 and W2 have been separately measured by us from the
FITS image with {\bf the {\tt AIPS} task `TVSTAT'}. Errors for E2 and W2 were assumed to 
be 7 per cent in both cases; hence, the error of on E2+W2 is the quadratic 
sum of the errors on E2 and W2.  \\ $^b$: The same procedure was followed 
as in $^a$; however, the errors for E2 and W2 were assumed to be 5 per cent. \\

{\bf References and comment}: The references are to the total flux
densities. \\ VLSS: VLA Low-frequency Sky Survey (Cohen et
al. 2007). We have measured the flux from the FITS file of the
map. 20 per cent error has been assumed for the 74 MHz integrated flux. \\ NVSS:
NRAO VLA Sky Survey (Condon et al. 1998). 5 per cent error in integrated flux
has been assumed. \\ 1: This work. Integrated flux values have been
measured by us directly from the maps with {\bf the {\tt AIPS} task `TVSTAT'}.  
2: Colla et al., 1970. Error has been assumed to be equal to the completeness 
limit of the survey.  3: Saikia et al., 2006. Flux values are measured by us
from the FITS maps .  4: White \& Becker, 1992. Error has been assumed
to be equal to the completeness limit of the survey.  5: The
integrated flux at 1400 MHz has been averaged. The errors in three
independent measurements have been propagated to calculate the error
in average integrated flux.  6: Griffith et al., 1990. The error has
been assumed to be equal to the completeness limit of the survey.  7: Becker, White
\& Edwards 1991. Error is 5.6 per cent as quoted by NED.  8: Gregory \& Condon,
1991. Error is 12.8 per cent as quoted by NED.  9: The integrated flux at
4850 MHz has been averaged. The errors in two independent measurements
have been propagated to calculate the error in average integrated
flux.  10: Since the outer lobes have errors as large as the fluxes at
4850 and 4830 MHz, we averaged the total outer lobe fluxes at 4850 and
4830 MHz to get a better flux value with less error at some average
frequency of 4840 MHz. The errors on the outer total fluxes at 4850
and 4830 MHz have been propagated to calculate the error at the
averaged frequency of 4840 MHz. For the spectrum of total outer lobes
this averaged point is considered instead of using the 4850 and 4830
MHz points separately.
\end{flushleft}
\end{table*}

\begin{table*}
\caption{The integrated flux densities, total flux densities of inner double and that of outer double, and the 
errors of these flux densities for J1835+6204 are listed in this table. Similar procedures as in Table~\ref{tab_flux_intinnout.j0041} have 
been followed to compile the flux densities. Column designations are self evident.
}
\label{tab_flux_intinnout.j1835}
\begin{tabular}{l l r r c r r c r r l}
\hline
Freq.    &Cp       & $S_{t}$ &  Error & Cp        & $S_{t}$ & Error & Cp        & $S_{t}$ & Error & Reference \\
 MHz     &         &  mJy    &  mJy   &           & mJy     & mJy   &           & mJy     & mJy   &   and      \\
         &         &         &        &           &         &       &           &         &       & comment   \\
(1)      &  (2)    &  (3)    & (4)    &  (5)      &  (6)    & (7)   & (8)       & (9)     & (10)  & (11)       \\
\hline
   38.00 & Int     &  19100  &  1910  &           &         &          &           &         &       &     1      \\  
   38.00 & Int     &  13700  &  1370  &           &         &          &           &         &       &     1      \\  
   38.00 & Int(avg)&  16400  &  1175  &           &         &          & SE1+NW1   &  11371  &  1772 &     2      \\  
  153.25 & Int     &  5510   &   826  &           &         &          & SE1+NW1   &   3994  &   850 &     3     \\
  240.25 & Int     &  5929   &   889  &           &         &          & SE1+NW1   &   4899  &   900 &     3      \\
  332.50 & Int     &  3375   &   506  &           &         &          & SE1+NW1   &   2596  &   517 &     3     \\
  326.00 & Int     &  2970   &   119  &           &         &          &           &         &       &     4 (WENSS)     \\  
  605.81 & Int     &  1813   &   127  &           &         &          & SE1+NW1   &   1348  &   143 &     3       \\ 
 1287.50 & Int-c   &   859   &    60  & SE2+NW2   &250$^\ast$&  13$^p$ & SE1+NW1   &    616  &    70 &     3       \\ 
 1400.00 & Int     &   800   &    40  &           &         &          & SE1+NW1   &    574  &    52 &     5 (NVSS)  \\ 
1442.50$^a$& Int-c &   708   &    35  & SE2+NW2   &218$^\ast$&   8$^q$ & SE1+NW1   &    488  &    48 &     6      \\ 
 4850.00 & Int     &   200   &    19  &           &         &          &           &         &       &     7      \\
 4850.00 & Int     &   195   &    29  &           &         &          &           &         &       &     8      \\
 4850.00 & Int(avg)&   198   &    17  &           &         &          & SE1+NW1   &    120  &    28 &     9      \\
4860.10$^b$& Int-c &   245   &    12  & SE2+NW2   &79$^\ast$&    3$^q$ & SE1+NW1   &    167  &    17 &    10     \\
8460.10$^c$& Int-c &   148   &     7  & SE2+NW2   &48$^\ast$&  1.7$^q$ & SE1+NW1   &    100  &    11 &    11      \\ 
\hline
\end{tabular}
\begin{flushleft}
(avg): The average flux density of the previous rows at the same frequency. \\ 
$\ast$: Power law spectra of total inner lobes (without the core) were constrained by least-squares fit with these data, 
and extrapolated to the lowest frequency. Flux density of total inner double (core subtracted) at all other frequencies was estimated from the fitted power law. \\

$^a$: From the archival VLA data set with project code, AL412A. \\
$^b$: From the archival VLA data set with project code, AM954. \\ 
$^c$: From the archival VLA data set with project code, AS586. \\

$^p$: Flux densities of SE2 and NW2 were separately measured by us from the FITS image with {\bf the {\tt AIPS} task `TVSTAT'}. 
Errors of SE2 and NW2 have been assumed to be 7 per cent apiece; hence, the error of SE2+NW2 is the quadratic sum of errors of E2 and W2. \\
$^q$: Same procedure has been followed as in $^q$; however, the errors
for SE2 and NW2 were assumed to be 5 per cent apiece. \\

{\bf Reference and comment}: The references are to the total flux densities. \\
WENSS: Westerbork Northern Sky Survey (Rengelink et al., 1997). 4 per cent error in total flux density has been assumed. \\ 
NVSS: NRAO VLA Sky Survey (Condon et al. 1998). 5 per cent error in integrated flux has been assumed. \\

1: Both Integrated flux density and error values are from 8C survey (Hales et al., 1995) as quoted by 
   Nasa Extragalactic Database (NED). 
2: This total flux density is the average of the two values of 38 MHz fluxes as quoted by NED.  
3: The integrated flux densities are from our GMRT measurements (project code: 10CKa01).
4: Westerbork Northern Sky Survey (WENSS, Rengelink et al., 1997).        
5: NRAO VLA Sky Survey (NVSS, Condon et al. 1998).            
6: The integrated flux density is from the data with project code, AL412A  and inner double flux density is from the data with project code, AS697.
7: The integrated flux density and error is from  Gregory \& Condon (1991). 
8: The integrated flux density and error is from  Becker, White \& Edwards (1991). 
9: The integrated flux density is the average of the above two rows. The error has been propagated from those in the above two rows.    
10: The integrated flux density is from the data with project code, AM954.  
11: The integrated flux density is from the data with project code, AS586. 
\end{flushleft}
\end{table*}

\subsection{Spectra}
\label{sec_spectra}
  The observational parameters and flux densities of the outer and
  inner lobes/doubles of J0041+3224 and J1835+6204 are listed in
  Table~\ref{obsparam.j0041} and \ref{obsparam.j1835}. Obviously, all
  the uv data used for imaging have different shortest baselines, and
  so we re-mapped the fields of both sources at high frequencies with
  a similar lower uv cutoff to image the inner double free from outer
  diffuse emission.  We tabulate the high-frequency flux densities and 
  their errors {\bf in Table~\ref{tab_inn.dbl.flux}} for the lobes and
  total inner doubles of J0041+3224 and J1835+6204, which were
  measured from the maps re-made with similar lower uv cutoff.  No
  appreciable curvature is visible in the spectra of the components
  (except the core) of the inner doubles of both the sources within
  our observable frequency range.  Therefore, we have fitted power
  laws to the spectral data of the inner lobes and the total inner
  doubles with the flux density and error values tabulated by Konar et
  al. (2012, in prep).  The flux density measurements of J0041+3224 at
  617, 1287, 4860 and 8460 MHz have been made from the FITS maps
  published by Saikia et al. (2006).  A power law has been fitted to
  only three high frequency points (namely, at 1287, 4860 and 8460
  MHz) tabulated {\bf in Table~\ref{tab_inn.dbl.flux}} to constrain the
  spectra of NW2 lobe, SE2 lobe and the total inner double of
  J0041+3224. These spectra have been extrapolated to lower
  frequencies.  Expressing $S$ in mJy and $\nu$ in MHz, the best fitting 
  power laws for the SE2 lobe and NW2 lobe of J0041+3224
  are $S_{SE2}(\nu) = (55865.9 \pm 14510)\times \nu^{-(0.698 \pm
    0.031)}$ and $S_{NW2}(\nu) = (45541.8 \pm 10940) \times
  \nu^{-(0.801 \pm 0.028)}$ respectively. The best fitting power law for
  the total inner double is $S_{inn} = (93737.3 \pm 23130)\times
  \nu^{-(0.724 \pm 0.030)}$.  For J1835+6204, we have fitted power
  laws to only four high frequency data points (namely, at 1287, 1440,
  4860 and 8460 MHz) tabulated {\bf in Table~\ref{tab_inn.dbl.flux}} to
  constrain the spectra of SE2 lobe, NW2 lobe and entire inner
  double. The best fitting power laws for SE2 and NW2 lobes are
  $S_{SE2}(\nu) = (81808.3 \pm 4859) \times \nu^{-(0.880\pm0.007)}$
  and $S_{NW2}(\nu) = (35066.7 \pm 8009) \times
  \nu^{-(0.830\pm0.028)}$ respectively.  The best fitting power law of the
  total inner double is $S_{inn}(\nu) = (114830.0 \pm 12800) \times
  \nu^{-(0.860\pm0.014)}$. For this source as well, we have used the same 
  units of $S$ and $\nu$ to express these power laws.  The fits of both 
  the sources along with the observed data points are plotted in
  Figure~\ref{spect_int.out.inn_j0041.n.j1835}.  The flux densities
of the integrated source, inner and outer doubles, are given in
Table~\ref{tab_flux_intinnout.j0041} and
\ref{tab_flux_intinnout.j1835}. The integrated spectra of the entire
source, and the spectra of the core-subtracted outer and inner doubles
are presented in Figure~\ref{spect_int.out.inn_j0041.n.j1835}.
Clearly there is steepening in the spectra of the integrated source
and the outer lobes at high frequencies.  The steepening of the
spectra at higher frequency is  presumably due to spectral ageing
(due to synchrotron and inverse-Compton losses). Low-frequency
measurements from our observations supplemented with those from the
literature show significant straightening of the spectra below
$\sim$300 MHz. This low frequency straight part of the spectra can be
characterised by a power law whose spectral index most likely
represents the injection spectral index ($\alpha_{\it inj}$). Further
analysis of these issues is described in
Section~\ref{spectral.ageing.analysis}.

\subsection{Radio core}
The J2000.0 positions of the radio core estimated from our high-resolution images are RA: $00\rm^h 41\rm^m 46\fs11$, DEC: $+32\degr 24^{\prime} 52\farcs83$ for 
J0041+3224 and RA: $18\rm^h 35\rm^m 10\fs41$, DEC: $+62\degr 04^{\prime} 07\farcs46$ for J1835+6204. For J0041+3224, a weak radio core has been detected only at
4860 MHz with a flux density of 0.57 mJy, and a weak peak of emission at 8460 MHz with a brightness of 0.2 mJy/beam 
at the position of the core has been found by Saikia et al. (2006).

The radio source J1835+6204 has a clear core detection at L band, C band and X band. The core flux densities as 
measured by us from our observations as well as from the VLA archival data have been compiled in Table~\ref{tab_core.flux_j1835}. 
Different observations were done at different epochs, enabling us to conclude that no prominent variability has been found
over a duration of 12 years.

\begin{table*}
\caption{Flux densities of the radio core of J1835+6204. 
The values of the core flux were estimated from two-dimensional Gaussian fits from our high resolution images.
The peak values from the fit are adopted as the core flux densities. Errors were calculated by quadratically adding 
the calibration errors and the {\tt JMFIT} (an {\tt AIPS} task) errors. The fiducial values of calibration errors are assumed to be 5 per cent for 
VLA data and 7 per cent for GMRT data. The column description of the table is as follows.
Columns 1, 2 and 3 show the telescope, the project code and the date of observations; column 4 shows the frequency 
of observations; column 5 lists the resolution of the map from which we measured the core flux density; and lastly, 
columns 6 and 7 list the core flux density and its error.      
}
\label{tab_core.flux_j1835}
\begin{tabular}{l c r l l l l}
\hline
Telescope  &  Project    &  Date of obs. & Freq. &  Resolution, PA                   &Flux      &Error    \\
           &  code       &               &       & $^{''}$$\times$$^{''}$, $^{\deg}$ &density   &         \\
           &             &               &  MHz  &                                    & mJy      & mJy     \\
  (1)      &   (2)       &  (3)          &  (4)  &      (5)                           &  (6)     & (7)        \\
\hline                                                                              
 GMRT      &  10CKa01    &  13-JUL-2006  &1287.5 & 02.10$\times$01.84, 338            &1.66      &  0.42   \\
 VLA-A     &  AS697      &  21-OCT-2000  &1440.5 & 01.35$\times$01.00, 357            &1.74      &  0.09    \\
 VLA-B     &  AL412A     &  18-FEB-1997  &1442.5 & 04.99$\times$03.66, 334            &1.62      &  0.18    \\
 VLA-B     &  AL412A     &  18-FEB-1997  &4860.1 & 01.52$\times$01.12, 324            &1.53      &  0.09    \\
 VLA-D     &  AM954      &  27-JUN-2008  &4860.1 & 15.31$\times$10.03,  46            &1.51      &  0.09   \\
 VLA-B     &  AL442.A    &  07-OCT-1998  &8460.1 & 00.79$\times$00.57, 336            &1.05      &  0.05   \\
 VLA-D     &  AS586.B    &  23-AUG-1996  &8460.1 & 08.79$\times$06.45,  14            &1.25      &  0.08   \\
\hline
\end{tabular}
\begin{flushleft}
\end{flushleft}
\label{core.flux.table}
\end{table*}

\section{Spectral ageing analysis}
\label{spectral.ageing.analysis}
In order to determine the spectral age in different parts of the
lobes, i.e. the time elapsed since the radiating particles were last
accelerated, we apply the standard theory describing the
time-evolution of the emission spectrum from particles with an initial
power-law energy distribution characterised by an injection spectral
index and distributed isotropically in pitch angle relative to the
magnetic field direction (JP model, see Jaffe \& Perola, 1973). Our
assumption is that the initial energy distribution over the entire
frequency/energy range of the spectrum of a blob of relativistic
plasma is characterised by the injection spectral index ($\alpha_{\rm
  inj}$). After sufficient amount of time is elapsed, the synchrotron
spectrum develops a curvature at high frequencies. This curvature is
characterised by a spectral break frequency. The spectral break
frequency ($\nu_{\rm br}$) above which the radio spectrum steepens
from the injected power law slope is related to the spectral age and
the magnetic field strength through
\begin{equation}
\tau_{\rm rad}=50.3\frac{B^{1/2}}{B^{2}+B^{2}_{\rm CMB}}\{\nu_{\rm br}(1+z)\}^{-1/2} {\rm Myr},
\label{eqn_specage}
\end{equation}

\noindent
where $B_{\rm CMB}$=0.318(1+$z$)$^{2}$ is the magnetic field strength equivalent to
the cosmic microwave background radiation; $B$ and $B_{\rm CMB}$ are expressed in units 
of nT, while $\nu_{\rm br}$ is in GHz. 

\subsection{Determination of $\alpha_{\rm inj}$ and $\nu_{\rm br}$ values}
There are three free parameters in the spectral ageing models: the normalisation, the injection spectral
index ($\alpha_{\rm inj}$) and the break frequency ($\nu_{\rm br}$). Determination of $\alpha_{\rm inj}$ requires 
very low frequency data points in the synchrotron spectrum.
\begin{figure*}
\hbox{
\vbox{
   \psfig{file=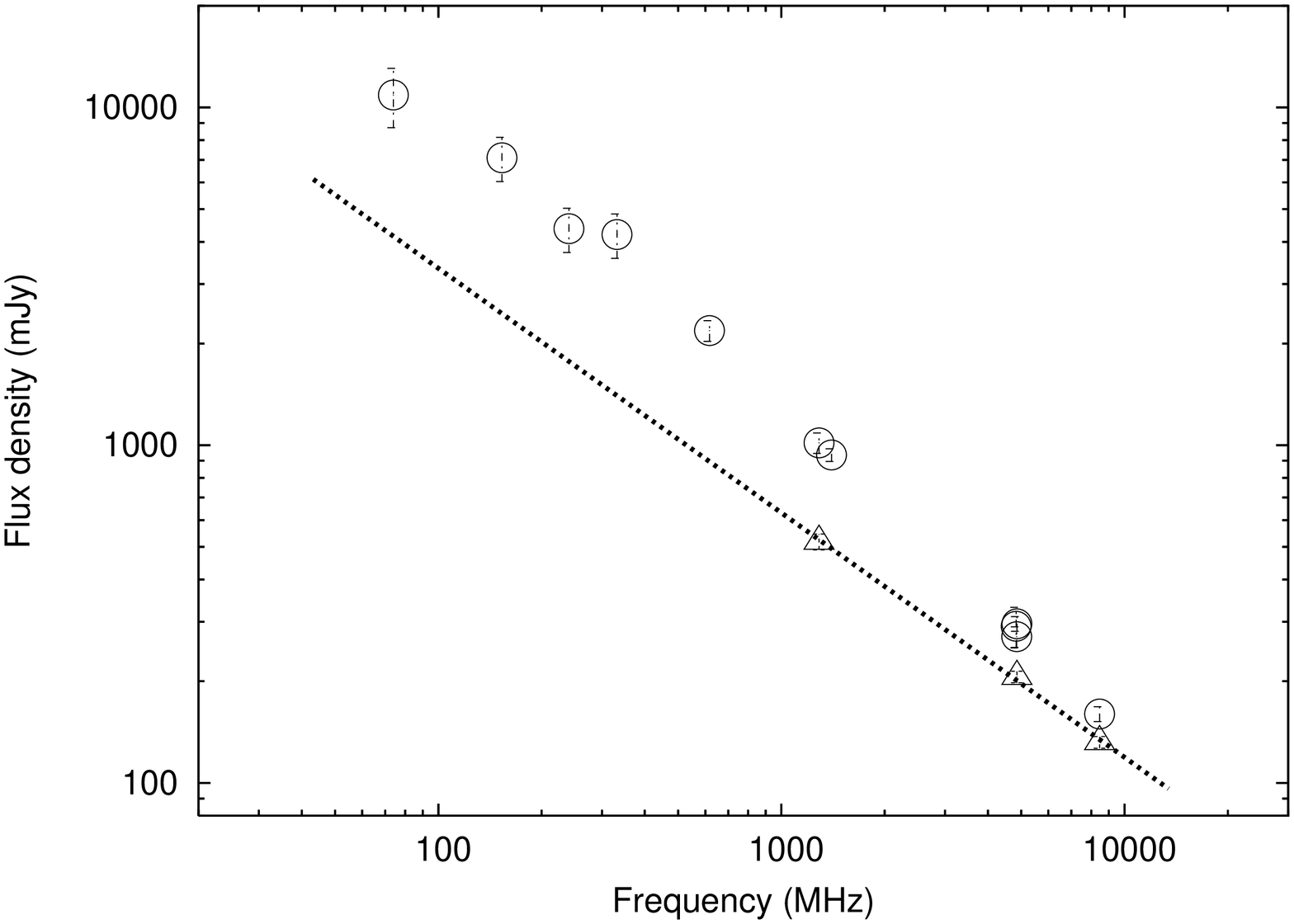,width=3.5in,angle=-90}
   \psfig{file=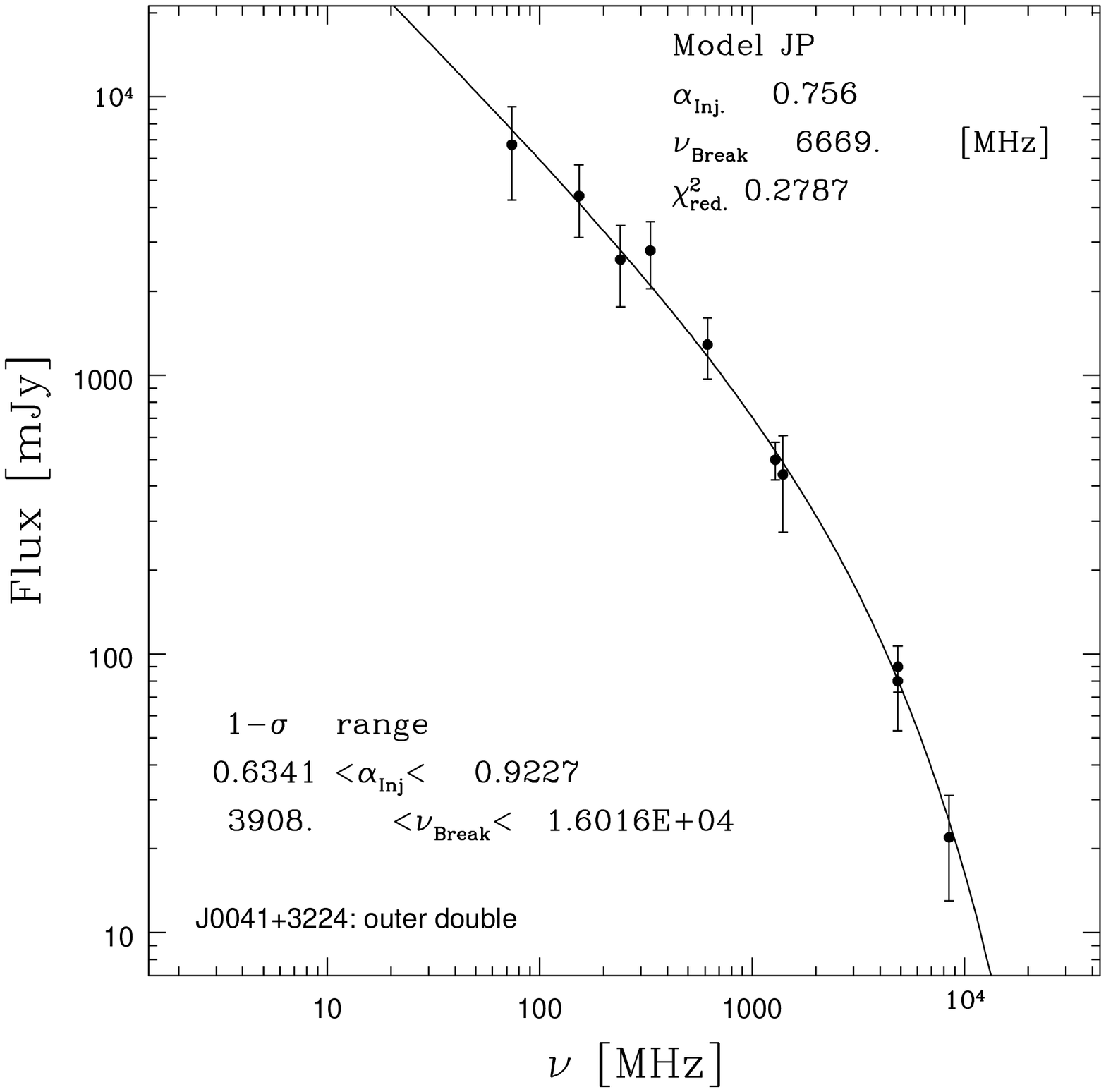,width=3.5in,angle=0}
}
\vbox{
   \psfig{file=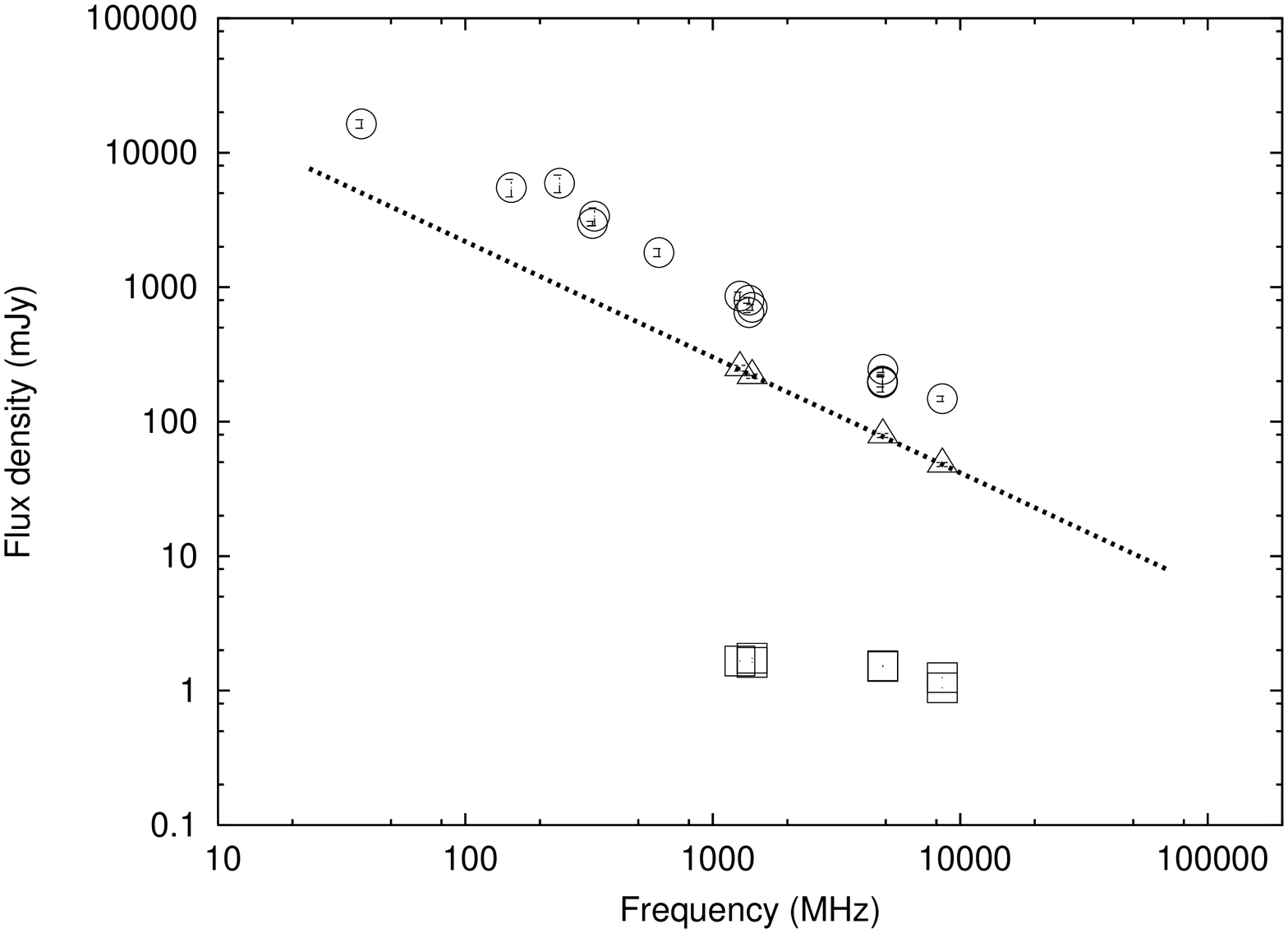,width=3.5in,angle=-90}
   \psfig{file=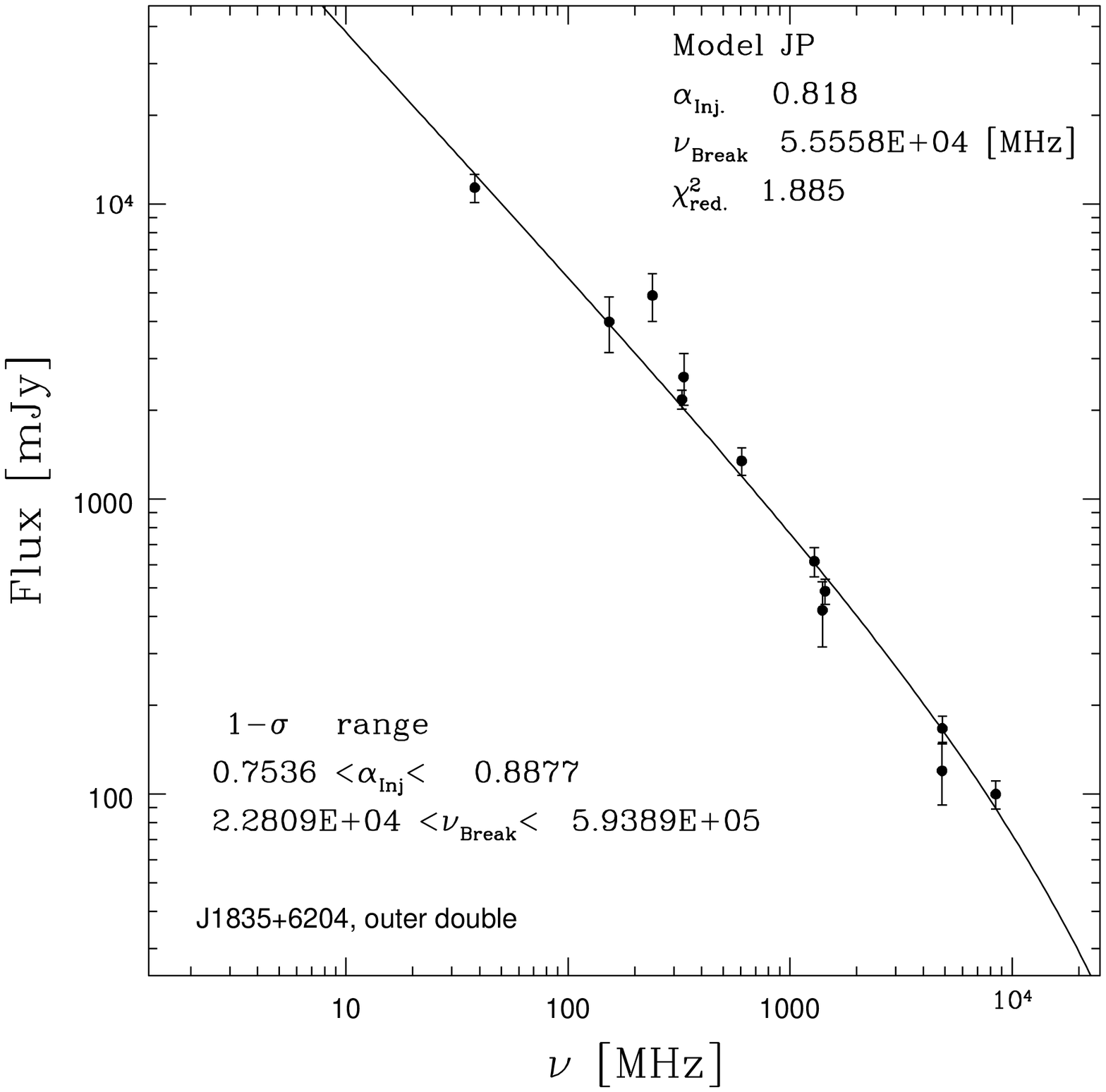,width=3.5in,angle=0}
}
}
\caption[]{Left upper panel: Integrated and inner double spectra of the source J0041+3224 have been plotted.
Open triangles are the inner double flux densities (core subtracted) measured from the high resolution images and
open circles are integrated flux densities. The inner double flux densities are fitted with a power law (dashed line) 
and extrapolated all the way back to 20 MHz to enable the reader to compare its strength with the integrated source. 
Left bottom panel: Filled circles are the flux densities of the entire outer double. The continuous curve represents 
the spectral ageing JP model fit.  
Right upper and right bottom panels are the same as the Left upper and Left bottom panels, respectively, but for the source
J1835+6204. The open squares in the upper right panel are the core flux densities (see Table~\ref{tab_core.flux_j1835}). 
}
\label{spect_int.out.inn_j0041.n.j1835}
\end{figure*}
High frequency data points are required for the accurate determination
of $\nu_{\rm br}$. The smaller the lower limit of the frequency span,
the better the determination of $\alpha_{\rm inj}$. Once $\alpha_{\rm
  inj}$ is determined accurately, the higher the upper limit of the
frequency span, the better the determination of $\nu_{\rm
  br}$. However, for high-frequency measurements, it is necessary to
be cautious about the loss of flux due to central uv holes in the
interferometric data, as well as the loss due to sparse uv coverage
and the low surface brightness of the lobes at higher frequencies.

Since low frequency flux measurements are very important in the
determination of $\alpha_{\rm inj}$, we have compiled very low
frequency measurements from the literature and tabulated in
Table~\ref{tab_flux_intinnout.j0041} and
\ref{tab_flux_intinnout.j1835}. While compiling the low frequency flux
densities from the literature, we have discarded those which are very
discrepant from the curve obtained by simple freehand drawing. A
simple estimate by eye clearly identifies data that are highly
discrepant. Since most of the measurements from the literature are
total flux density values, and no flux density values exist for a
single lobe component of a source, we have constrained the total
synchrotron spectrum of the two outer lobes together for each source
to very low frequencies.  The flux densities and their errors
  compiled in Tables~\ref{tab_flux_intinnout.j0041} and
  \ref{tab_flux_intinnout.j1835} have been used in fitting the spectra
  of various components of J0041+3224 and J1835+6204. The flux
  densities of the total inner doubles of both the DDRGs are fitted
  well with power laws and these best fitting power laws are given in
  Section~\ref{sec_spectra}. We have estimated the flux densities of the
  inner double from the best-fitting power law models at those
  frequencies for which we could not directly measure the flux
  densities of the inner doubles due to contamination of diffuse
  emission of outer double and/or poor resolution.  For those
  estimated flux densities the errors have been calculated using the 
  formula $\sigma_{S_{\nu}}=\sqrt\{(\frac{S_{\nu}}{S_0})^2 \sigma_{S_0}^2
  + (S_{\nu}ln\nu)^2 \sigma_{\alpha}^2 \}$, where $S_0$ (the
  normalisation), $\sigma_{S_0}$ (the error in normalisation) and
  $S_{\nu}$ are in mJy, $\nu$ is in MHz and $\sigma_{\alpha}$ is the 
  error in spectral index of a general power law spectrum of the form 
  $\frac{S_{\nu}}{mJy}=S_0(\frac{\nu}{MHz})^{\alpha}$.

In order to determine a value of $\alpha_{\rm inj}$, we first fitted
the JP (Jaffe \& Perola 1973) and Continuous Injection (CI) models
(Pacholczyk, 1970) of radiative losses to the flux densities of the
outer lobes. We found that the CI model does not provide a good fit
to the data. The application of the CI model to the outer structure is
physically unjustifiable for DDRGs in general, because a) the jets no
longer feed the outer lobes, and b) the outer lobes are so old that
they have suffered synchrotron and inverse-Compton losses to a large
extent (and possibly adiabatic loss to some extent). So high energy
electrons are no longer being replenished in the outer lobes.
Therefore, it is reasonable that the CI model does not provide a good
fit for the total spectrum of the outer lobes, and instead the JP
model is more appropriate. However, for J1835+6204, the NW1 lobe has a
clear compact hotspot (a signature of the ongoing feeding by the last
ejected jet material in the previous episode of JFA) and both the NW1
and SE1 lobes have single power-law spectra within our observable
range of radio frequencies.  Even in that case the CI model does not
yield a good fit, while the JP model does. This might have
something to do with the fact that most of the parts of the lobes
except the hotspot are old and the high energy particles are not
replenished. Since we cannot constrain the radio spectra of each lobe
separately up to very low frequencies, we have assumed that
$\alpha_{\rm inj}$ is the same for both the outer lobes, and that
$\alpha_{\rm inj}$ remains constant over the entire active phase of
JFA. 
\begin{figure*}
\vbox{
    \hbox{
   \psfig{file=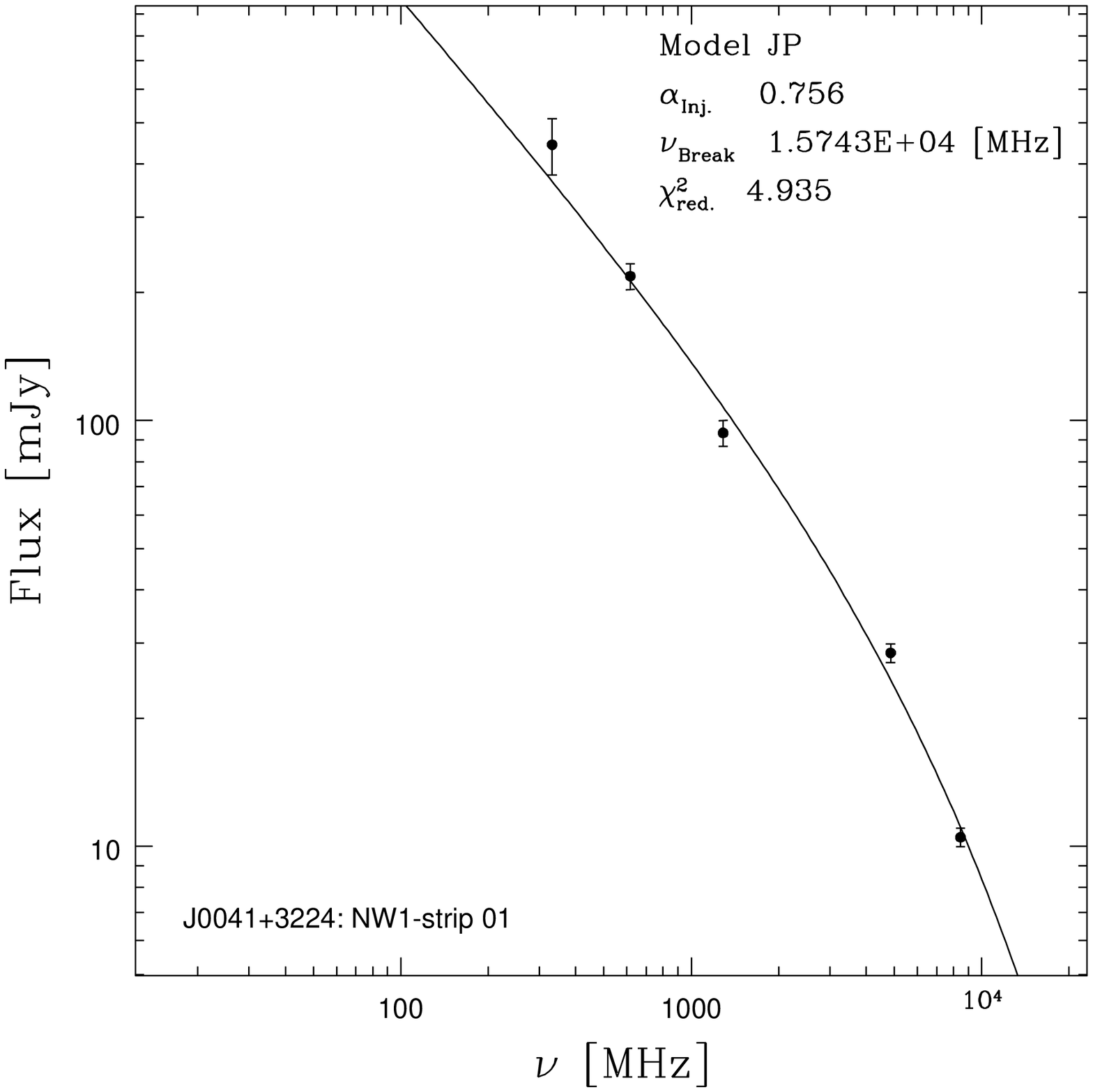,width=1.75in,angle=0}
   \psfig{file=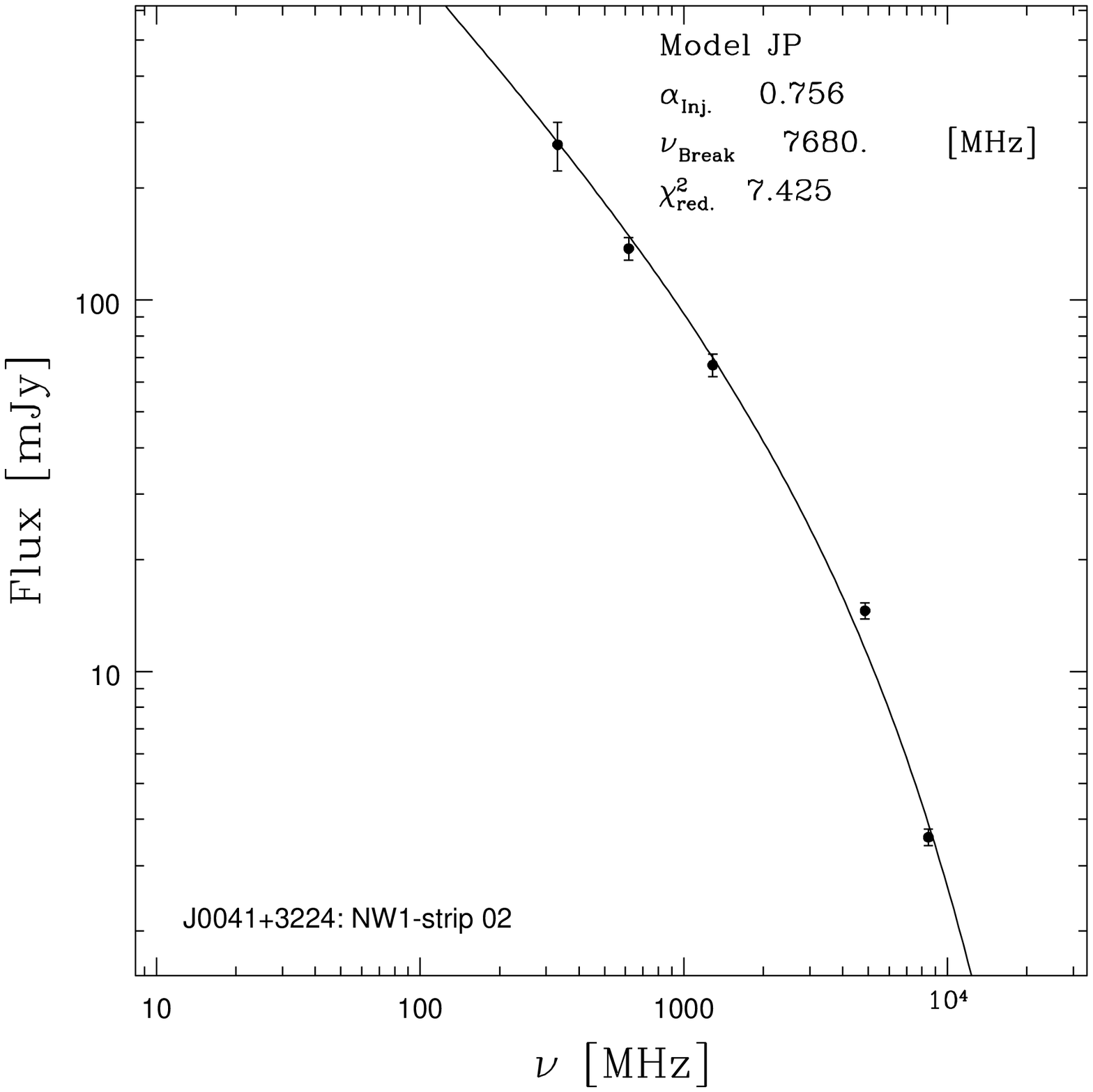,width=1.75in,angle=0}
   \psfig{file=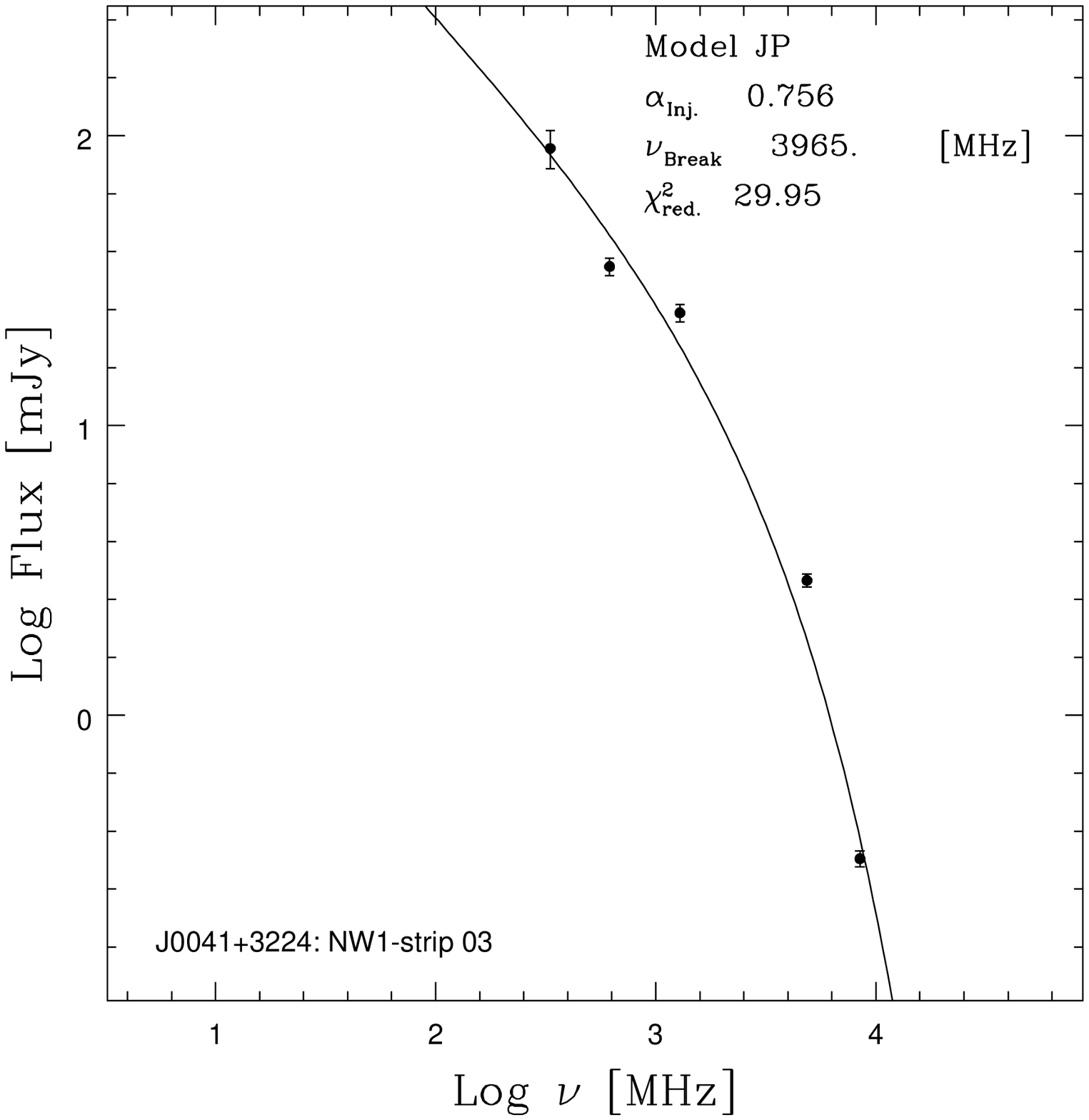,width=1.75in,angle=0}
   \psfig{file=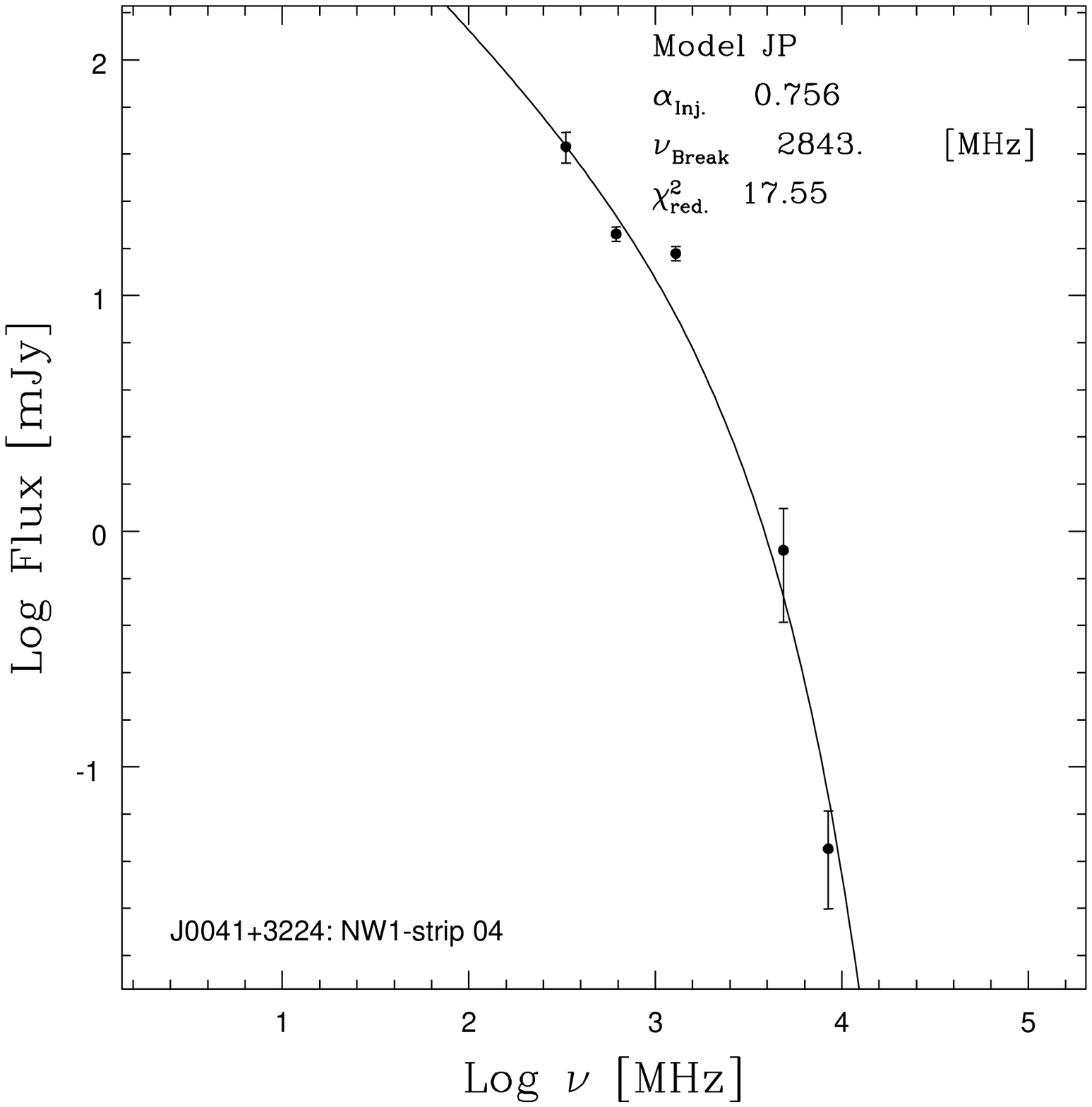,width=1.75in,angle=0}
   }
   \hbox{
\psfig{file=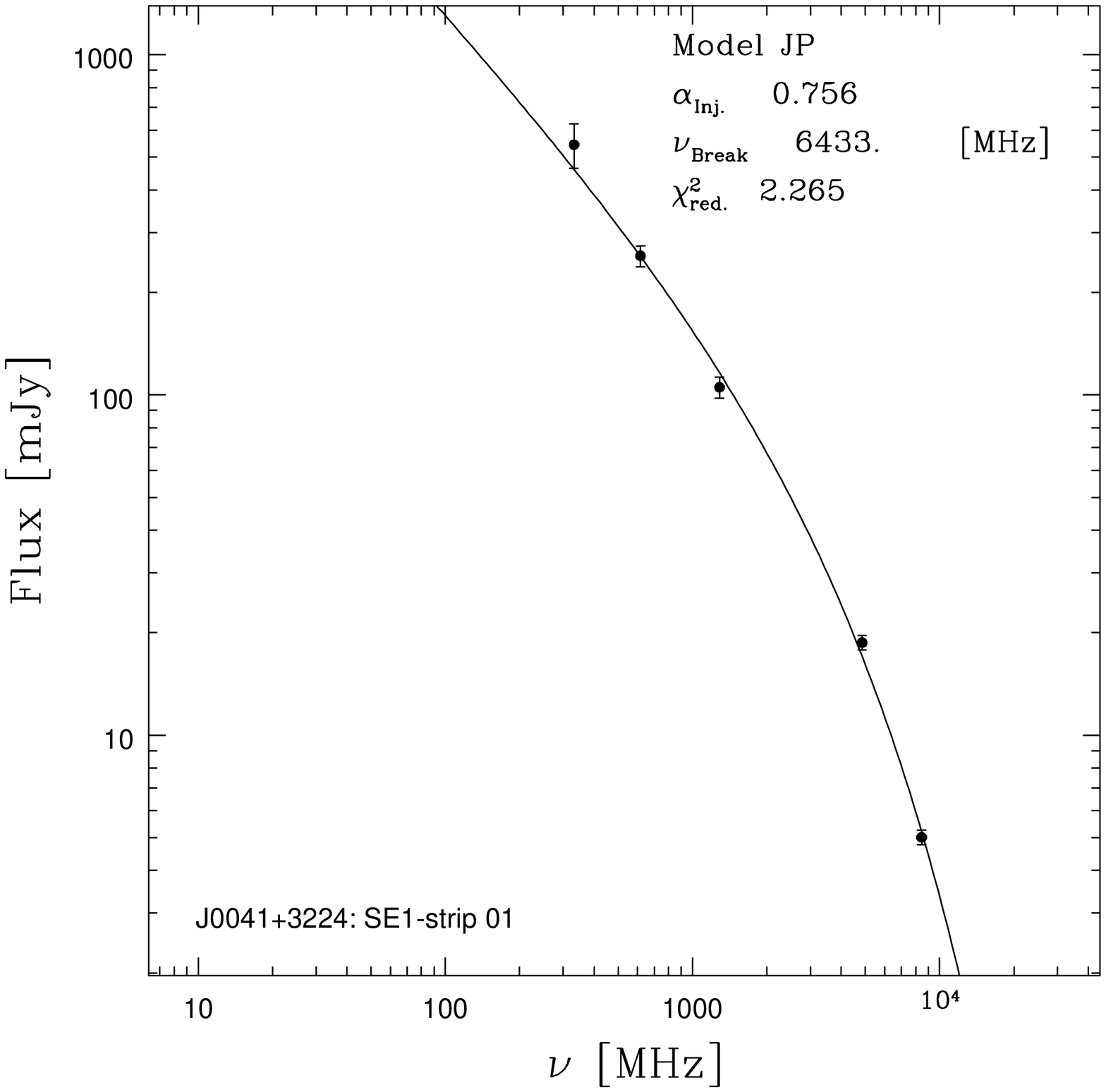,width=1.75in,angle=0}
\psfig{file=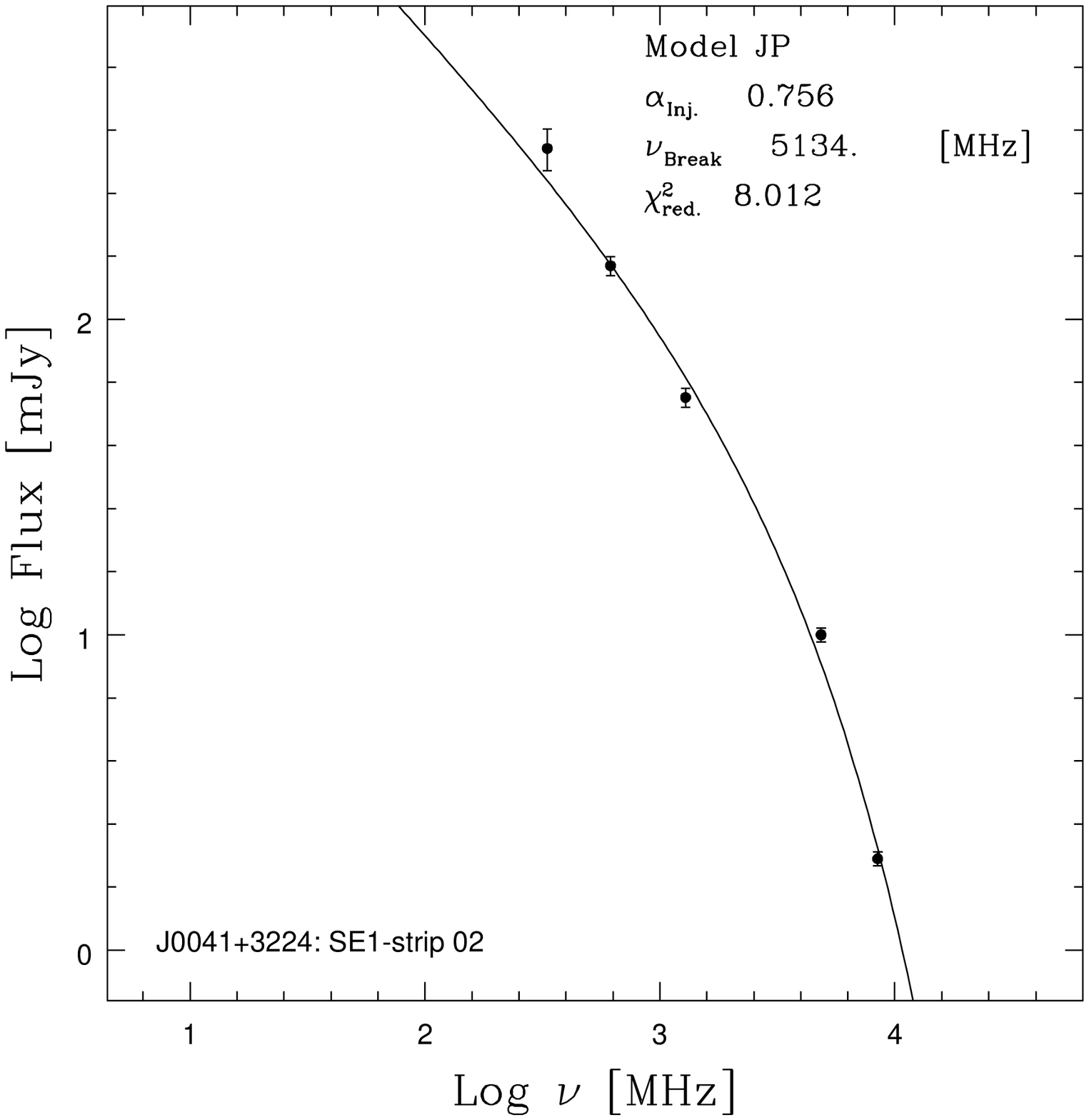,width=1.75in,angle=0}
\psfig{file=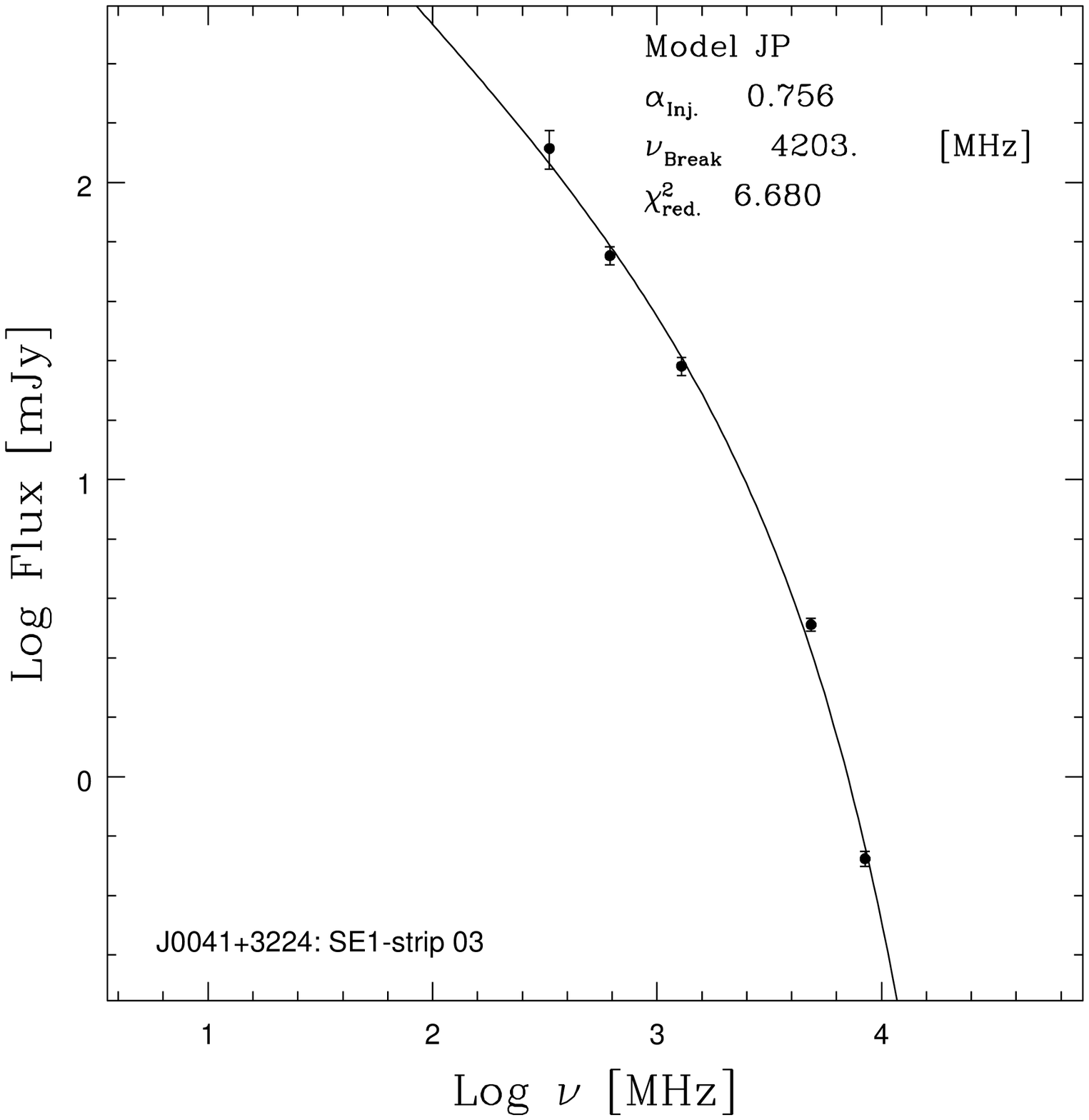,width=1.75in,angle=0}
\psfig{file=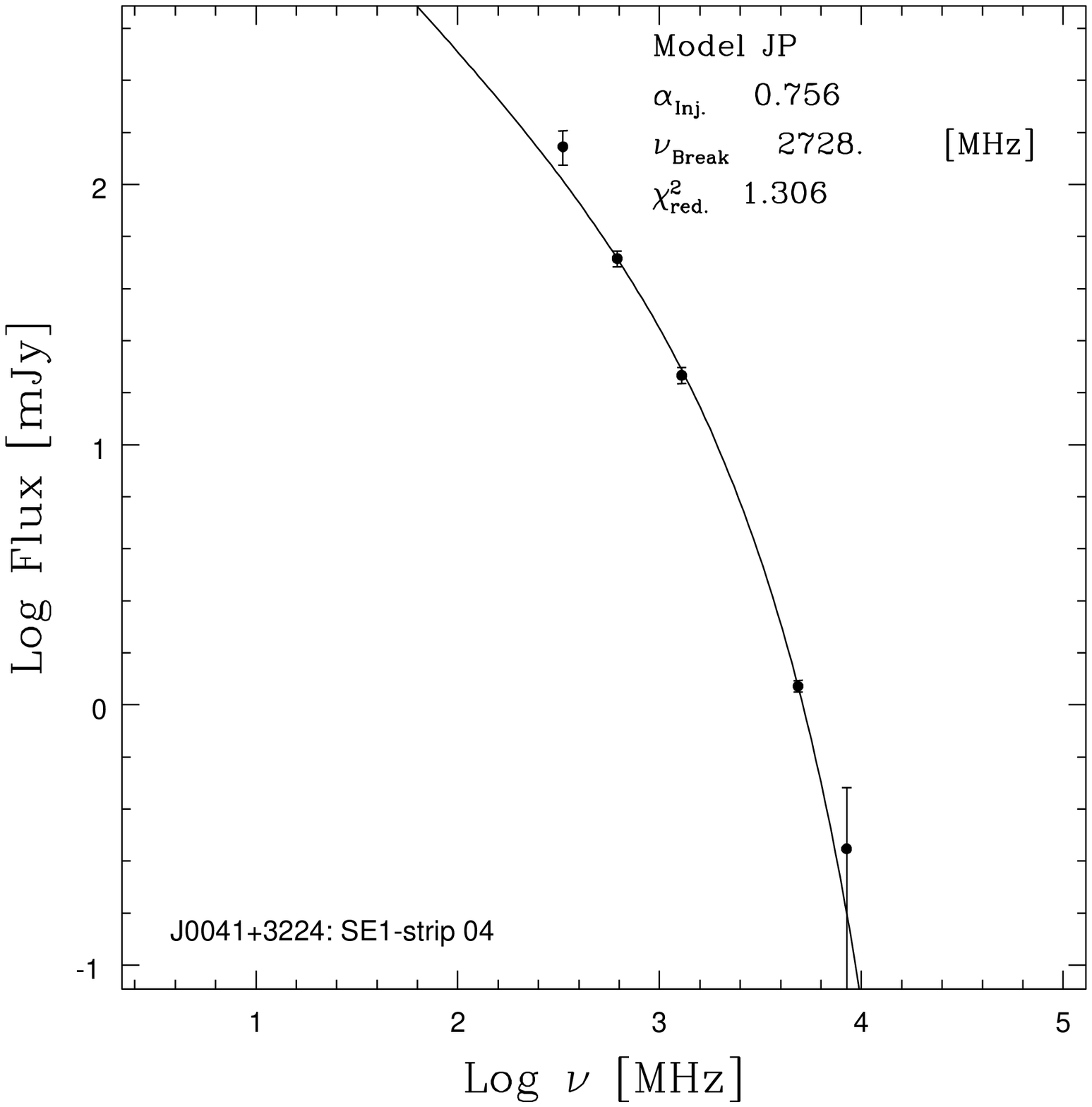,width=1.75in,angle=0}
    }
}
\caption[]{Typical spectra of a few slices for the eastern (upper panel) and western (lower panel)
lobes of the outer double of J0041+3224.}
\label{strip.fit_j0041}
\end{figure*}
\begin{figure*}
\vbox{
   \hbox{
   \psfig{file=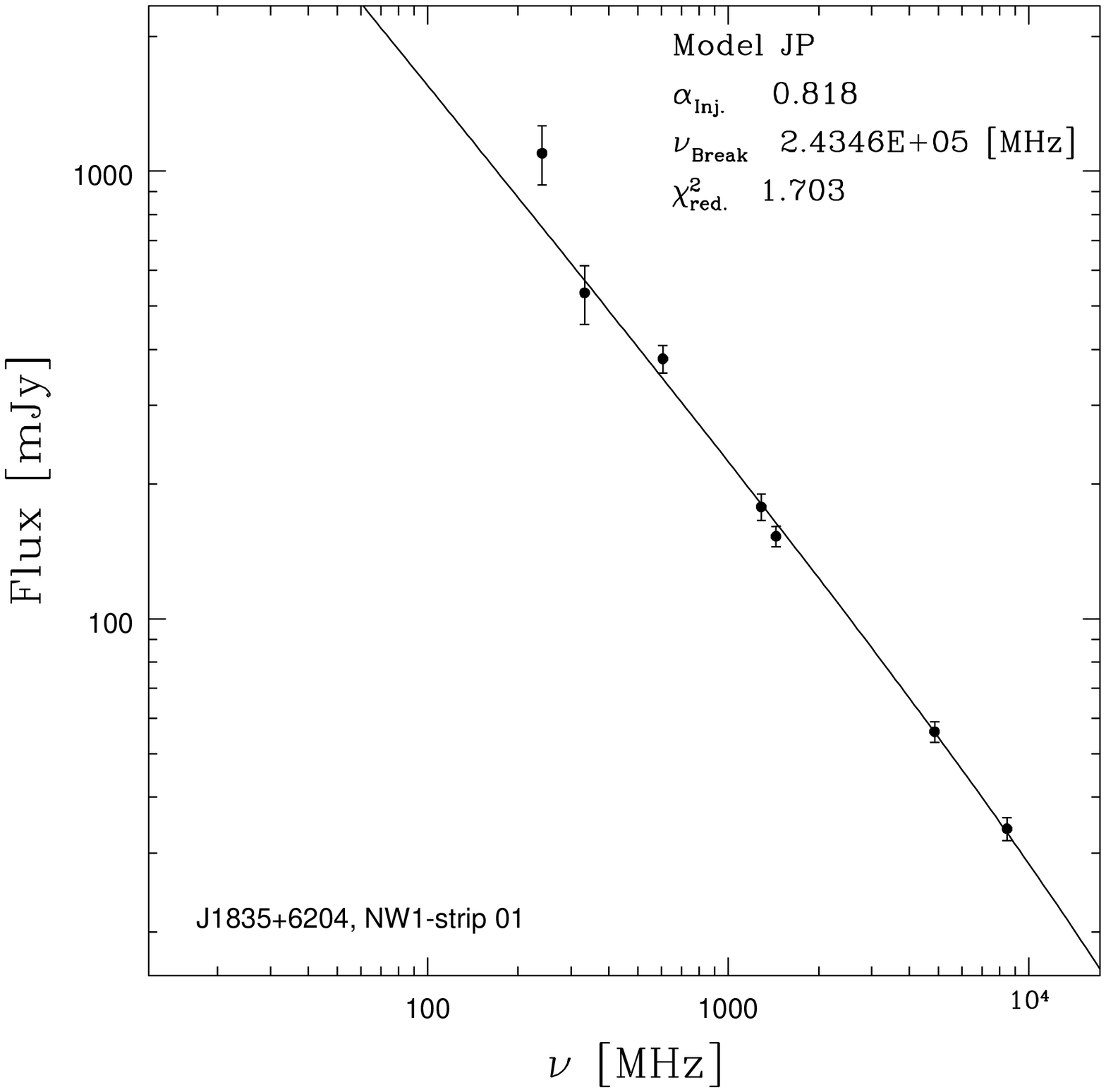,width=1.75in,angle=0}
   \psfig{file=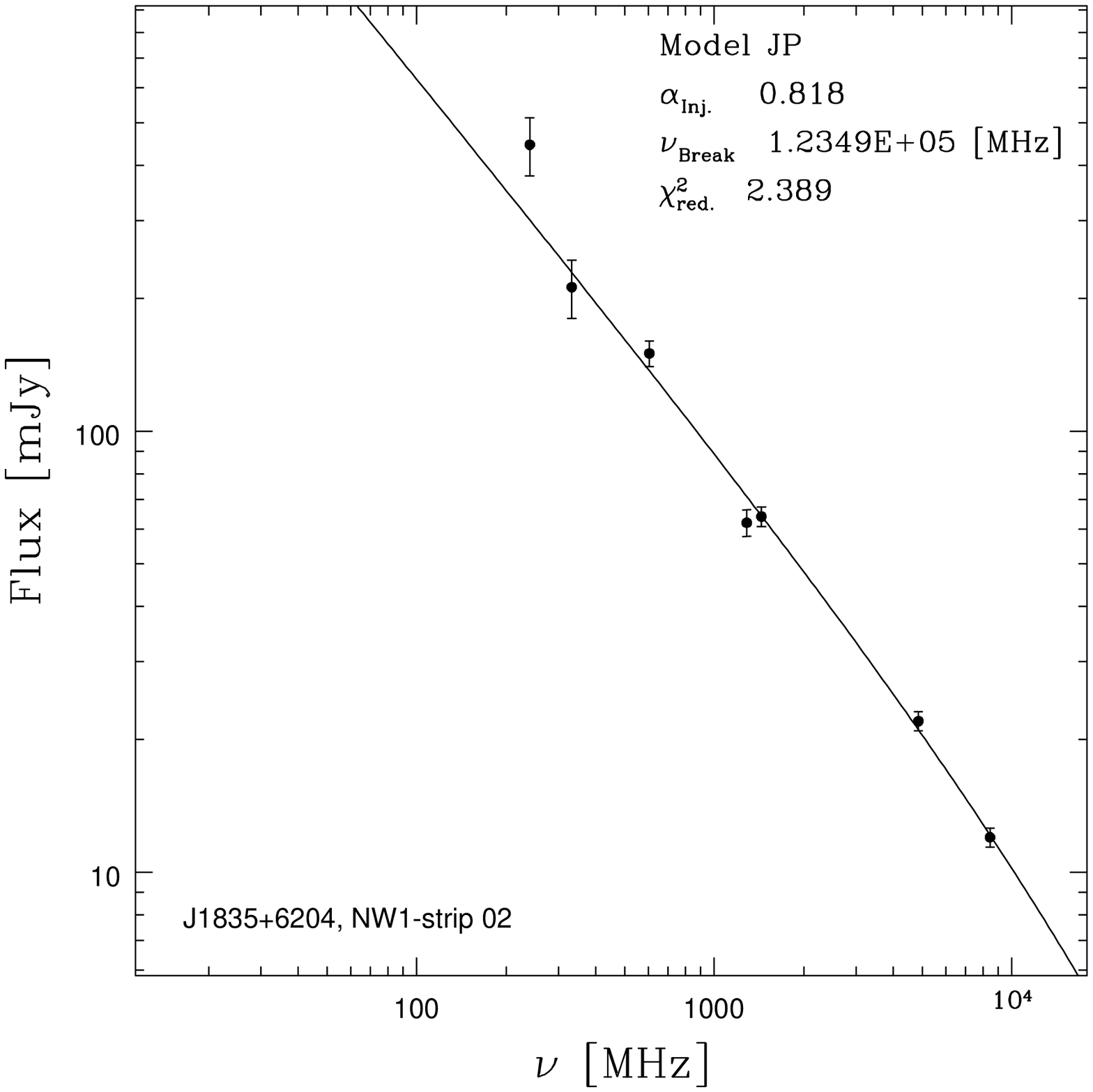,width=1.75in,angle=0}
   \psfig{file=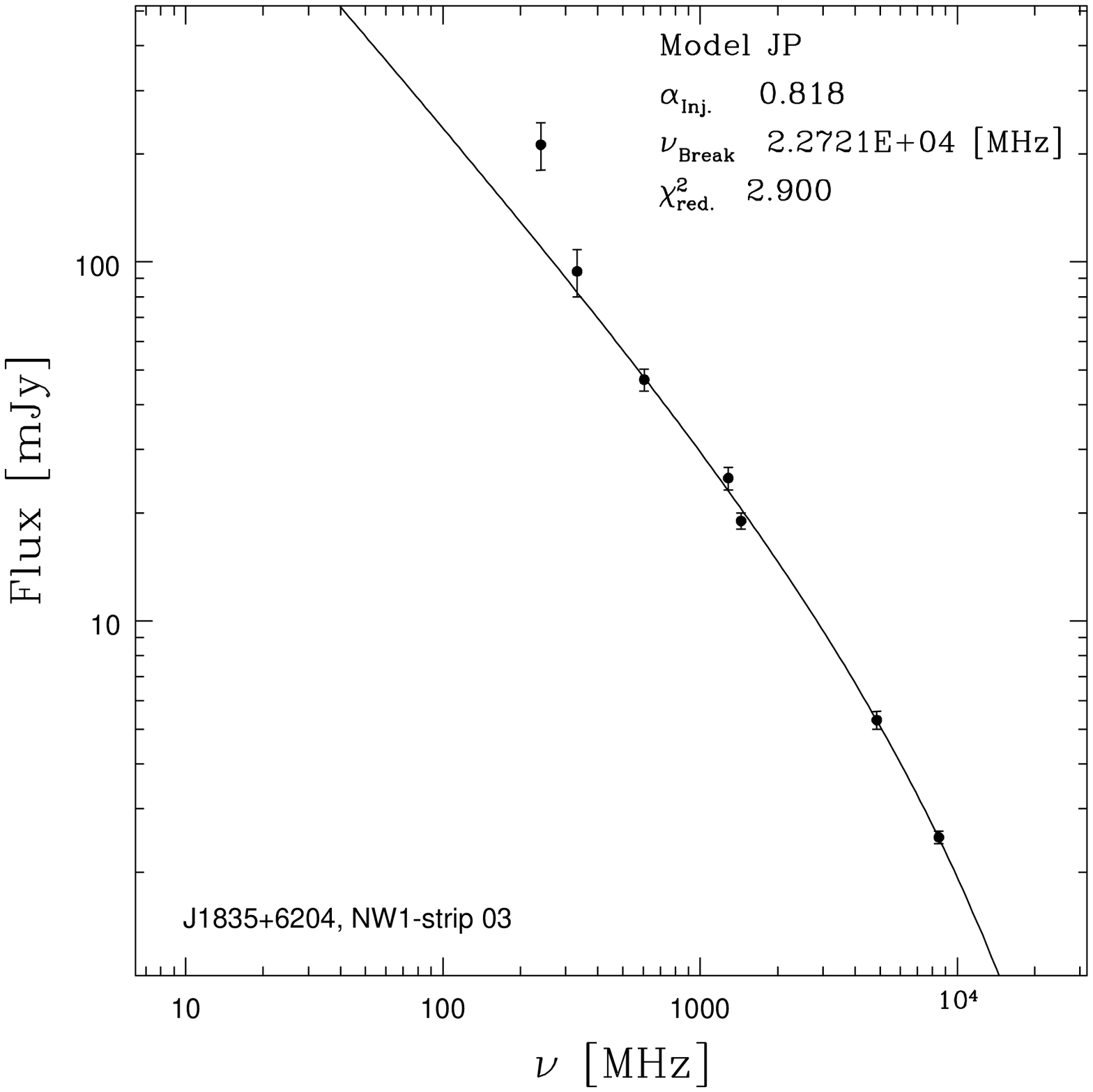,width=1.75in,angle=0}
   \psfig{file=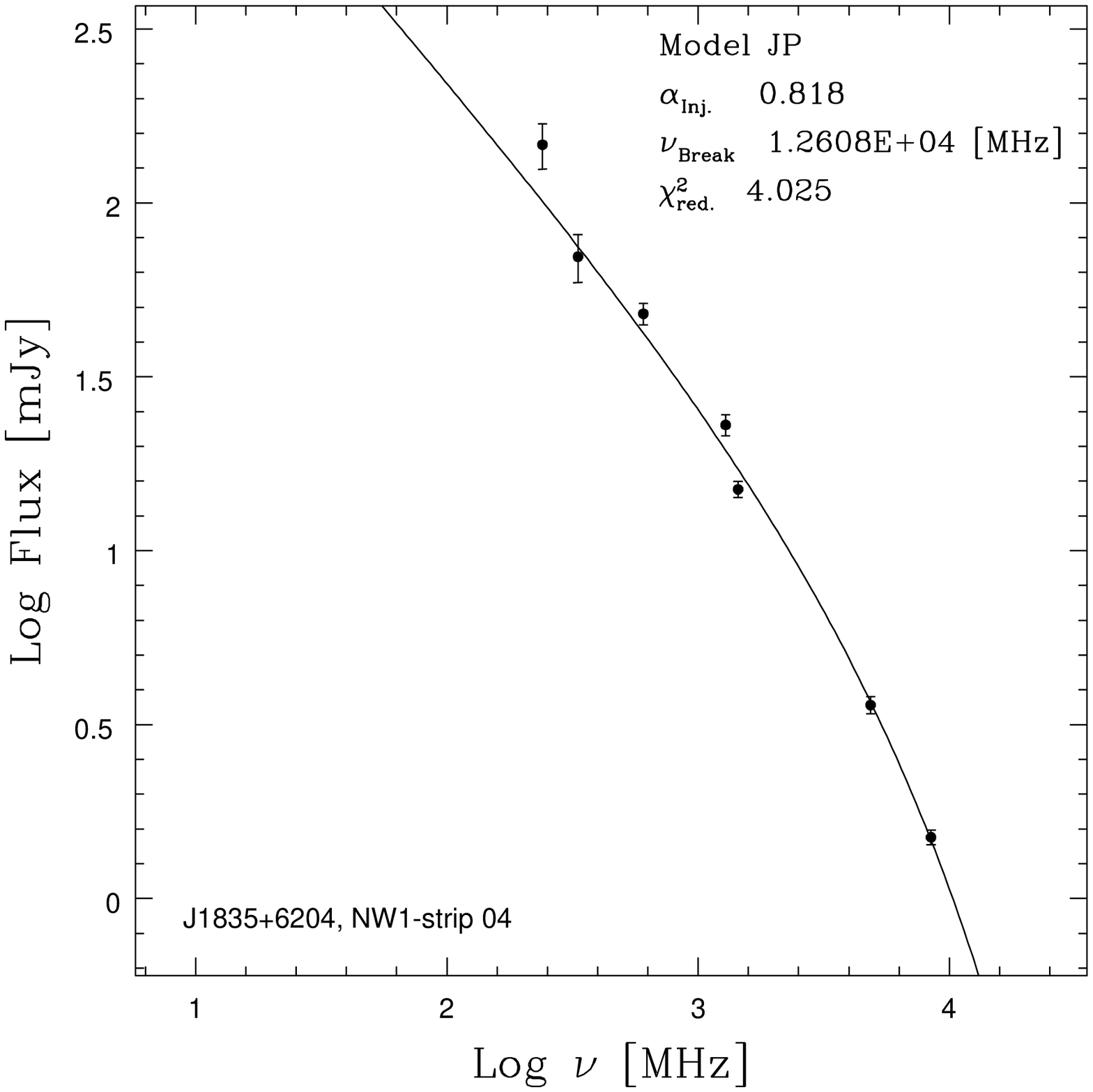,width=1.75in,angle=0}
    }
    \hbox{
   \psfig{file=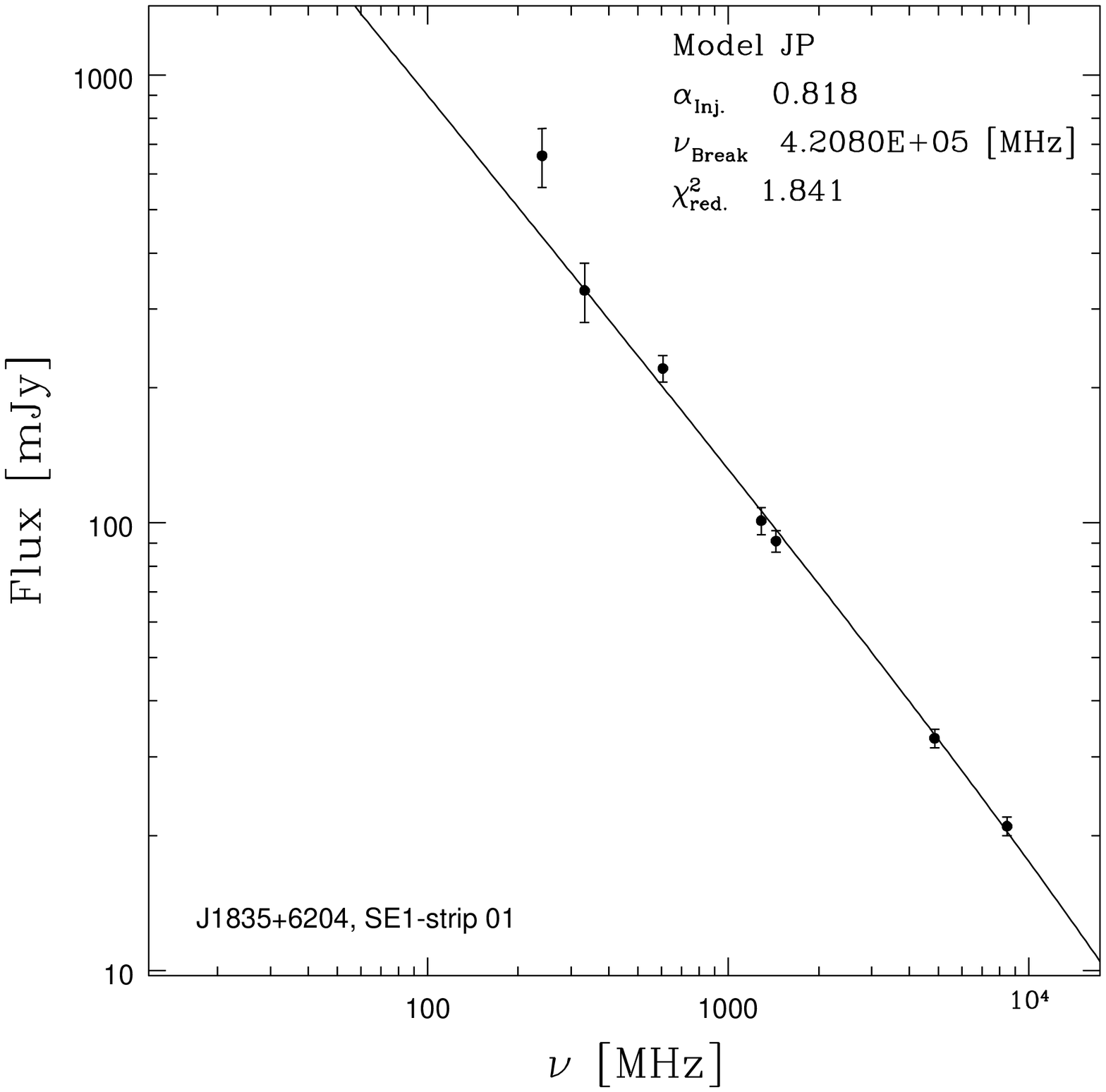,width=1.75in,angle=0}
   \psfig{file=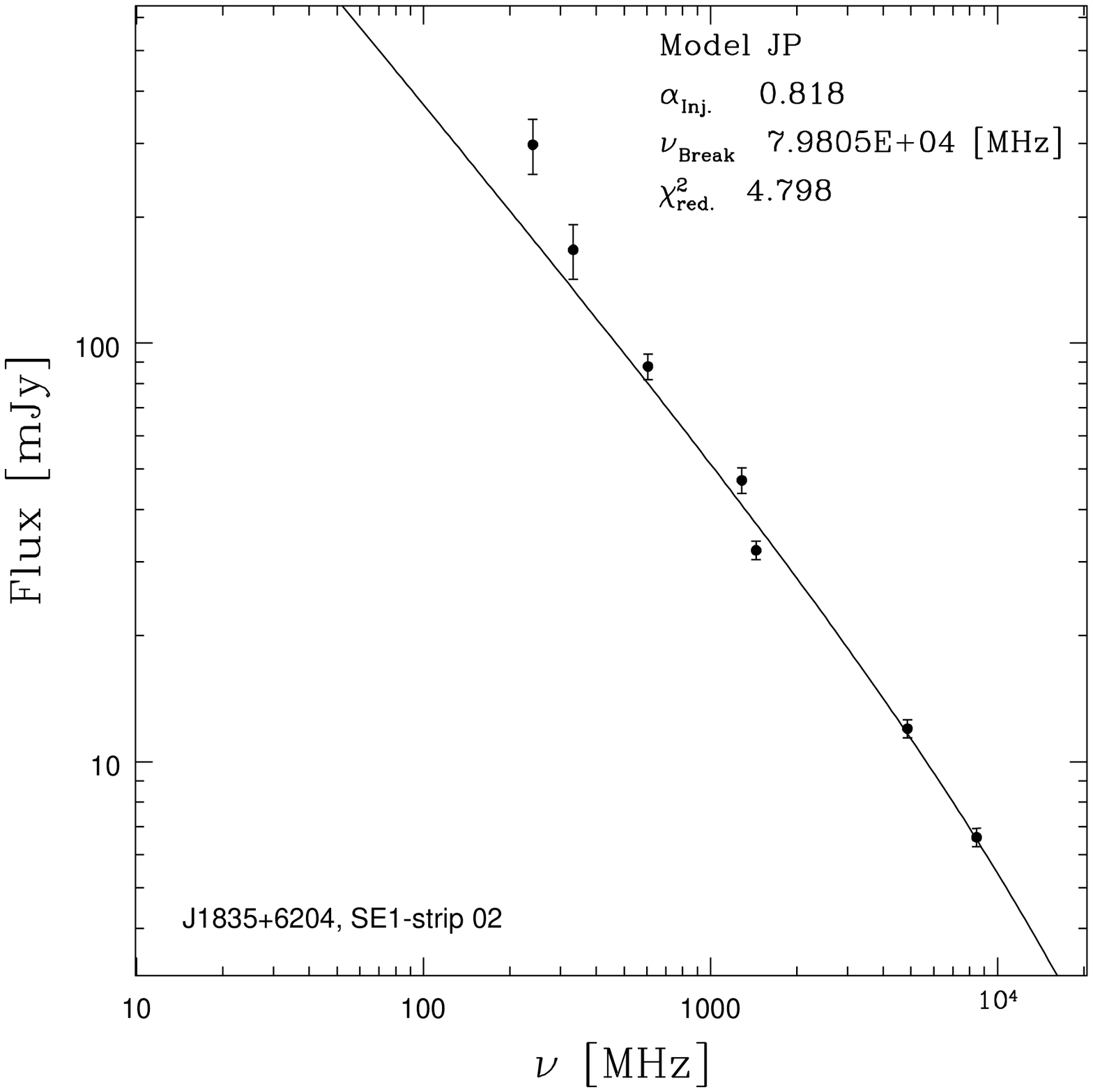,width=1.75in,angle=0}
   \psfig{file=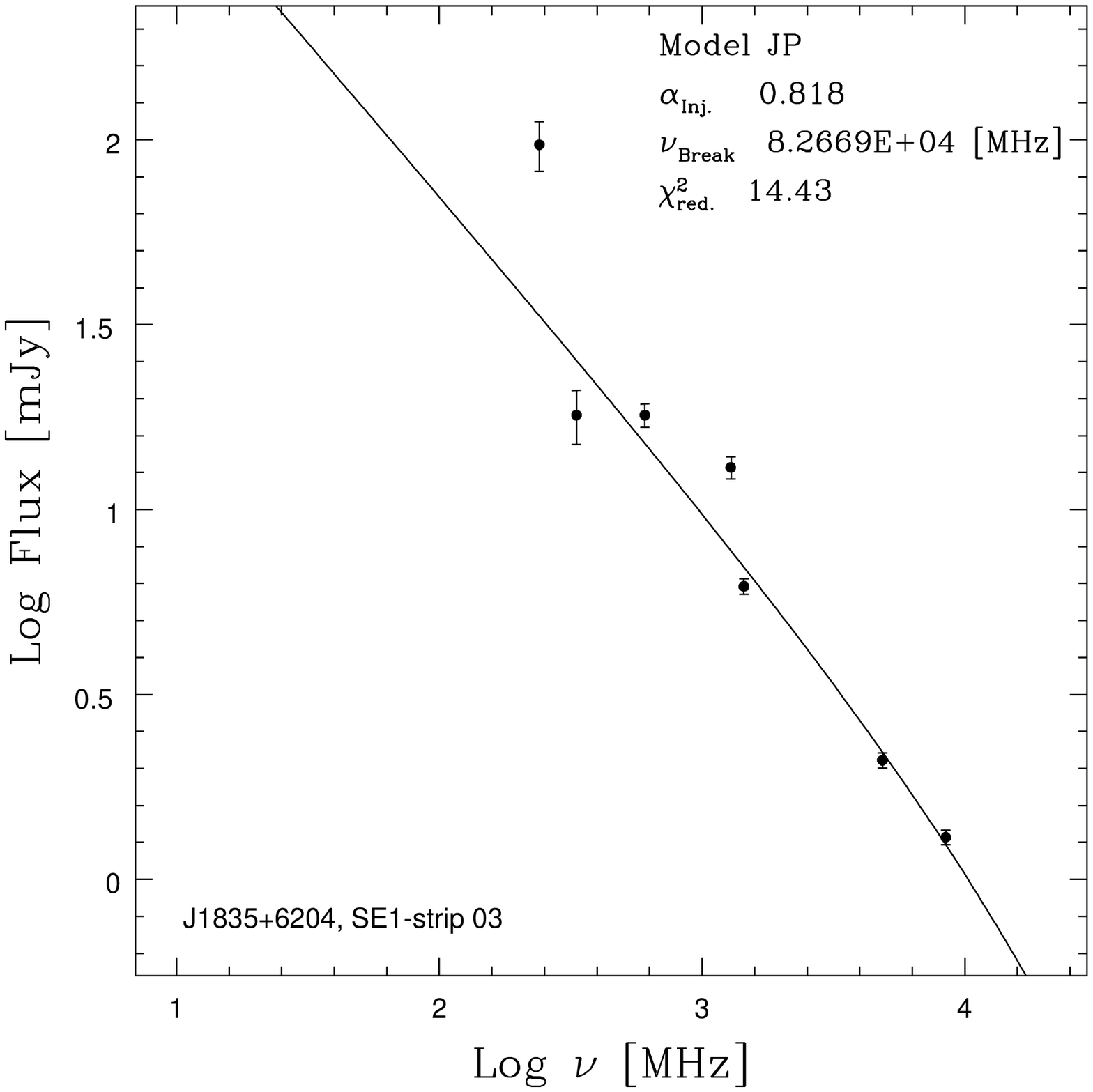,width=1.75in,angle=0}
   \psfig{file=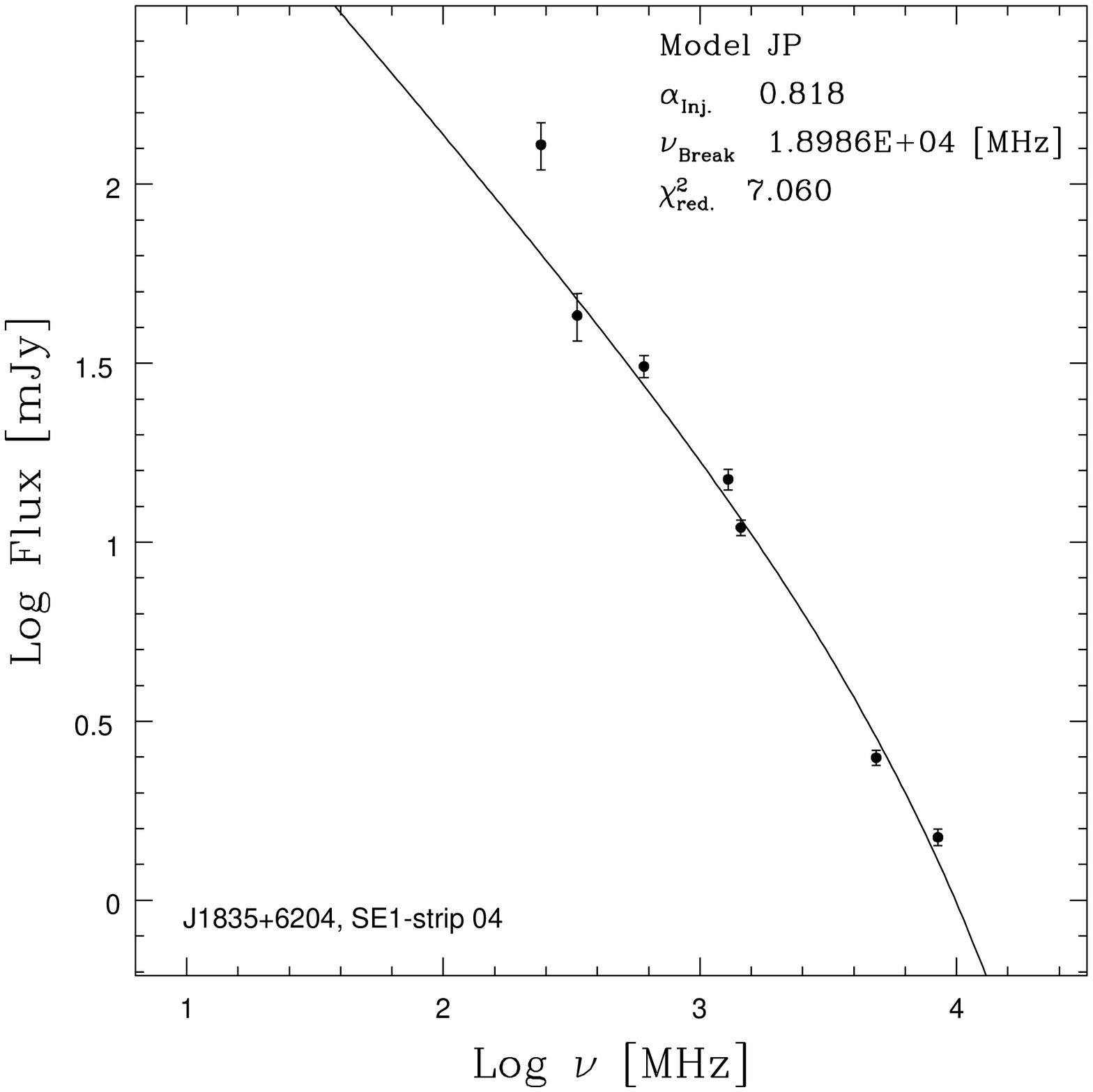,width=1.75in,angle=0}
   }
}
\caption[]{Typical spectra of a few slices for the north-western (upper panel) and south-eastern (lower panel)
lobes of the outer double of J1835+6204.}
\label{strip.fit_j1835}
\end{figure*}

The various best fitting parameters of the spectra of the outer lobes of
our 2 DDRGs are shown in Figure~\ref{spect_int.out.inn_j0041.n.j1835}
  These fits have
been used to constrain the values of $\alpha_{\rm inj}$ for the ageing
analysis of the outer lobes of the sources. Since there is no visual
indication of curvature in the spectra of the inner doubles, we have
fitted a pure power law to the inner lobes of both the DDRGs.  Then,
the total-intensity maps made by us at multiple frequencies were
convolved with a common angular resolution (circular beams of
$9\farcs50$ for J0041+3224 and $16\farcs70$ for J1835+6204).  Each
lobe was then split into a number of strips, separated approximately
by the common resolution element (with which the maps were convolved)
along the axis of the source, and the spectrum of each strip has been
constrained by fitting a spectral ageing model with our measured flux
densities. For fitting the synchrotron ageing model to the observed
spectra of our sources, we have used the {\tt SYNAGE} software (Murgia
1996), which is specifically written for this purpose. {\tt SYNAGE}
was used to fit a JP model to the spectrum of each strip of the outer
lobes of our sample sources. For a given source, we fixed the value of
$\alpha_{\it inj}$, as determined by our fits, shown in
Figure~\ref{spect_int.out.inn_j0041.n.j1835}, for all strips. The
fitted spectra of some of the strips are shown in
Figure~\ref{strip.fit_j0041} and \ref{strip.fit_j1835}. If we treat
$\alpha_{\rm inj}$ as a free parameter, then the resulting best-fitting
values of $\alpha_{\rm inj}$ for different strips show significant
variation with large error bars.  We believe that the large errors of
$\alpha_{\rm inj}$ in the fit are possibly due to the large
uncertainty in the surface brightness for each strip. So, for all the
strips of both lobes of each source, we have fixed the value of
$\alpha_{\rm inj}$ to the best fitting value obtained from the JP model
fit to the observed spectrum of entire outer double. The values of
$\alpha_{\rm inj}$ used for all the strips (of a source) are quoted in
the caption of Table~\ref{age.strips_j0041} and \ref{age.strips_j1835}
for J0041+3224 and J1835+6204 respectively.  Here our assumption is
that the jet injected all the relativistic plasma with the same value
of $\alpha_{\rm inj}$ over the entire period of the active phase. We
would need much more sensitive data to treat $\alpha_{\rm inj}$ as
free parameter for the spectrum of each strip. The values of $\nu_{\rm
  br}$ including the 1$\sigma$ errors for the strips of outer lobes
are listed in Table~\ref{age.strips_j0041} and
\ref{age.strips_j1835}. In our previous work, whenever we could
constrain $\alpha_{\rm inj}$ for each outer lobe separately, we found
that they are not very different (Jamrozy et al. 2008, Machalski et
al., 2010), and so there is no strong evidence for significantly
different values of this parameter in two opposite lobes of a radio
galaxy.

\subsection{Magnetic field determination and radiative ages}
\begin{figure}
\vbox{
    \psfig{file=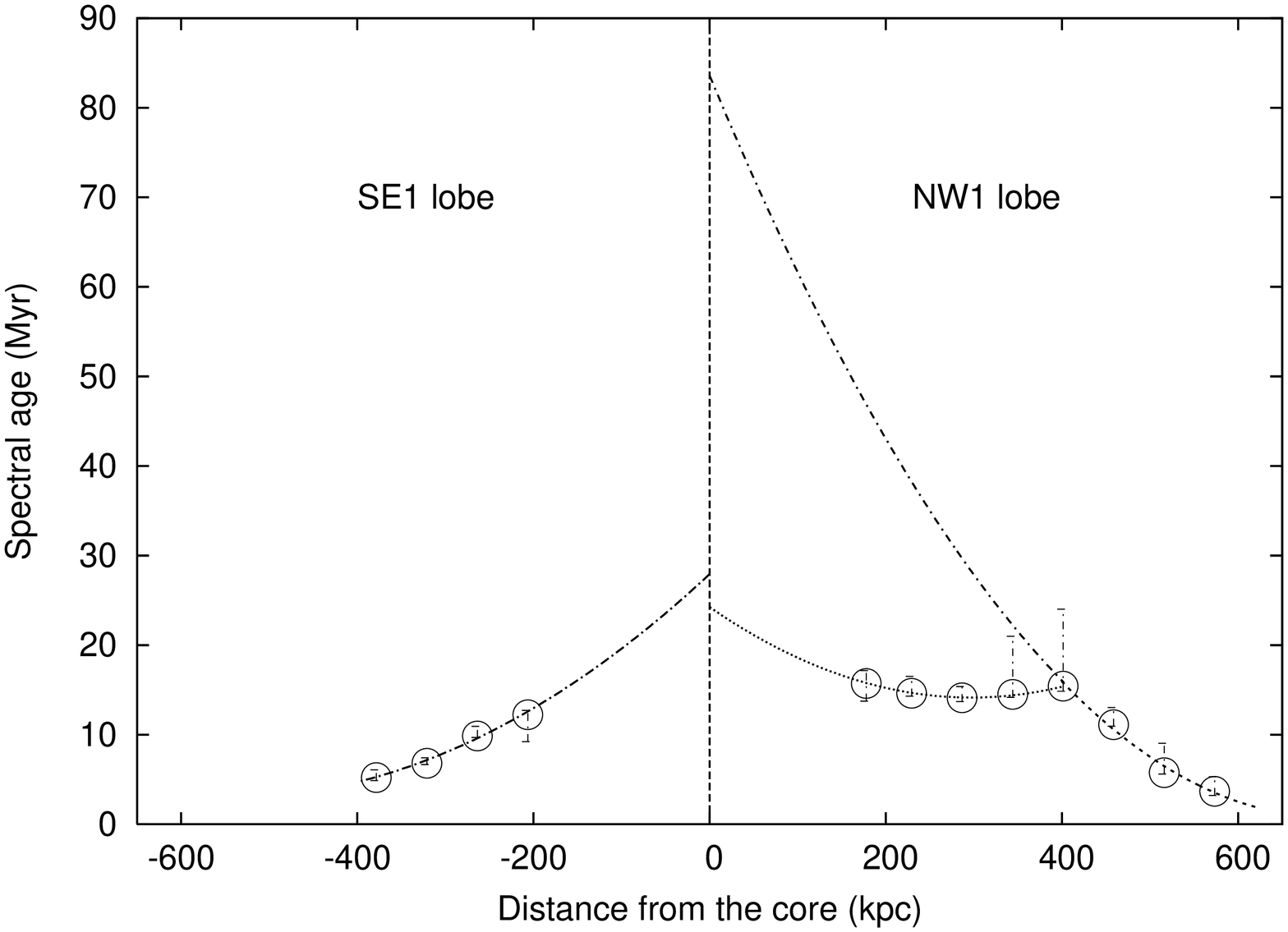,width=3.0in,angle=-90}
    \psfig{file=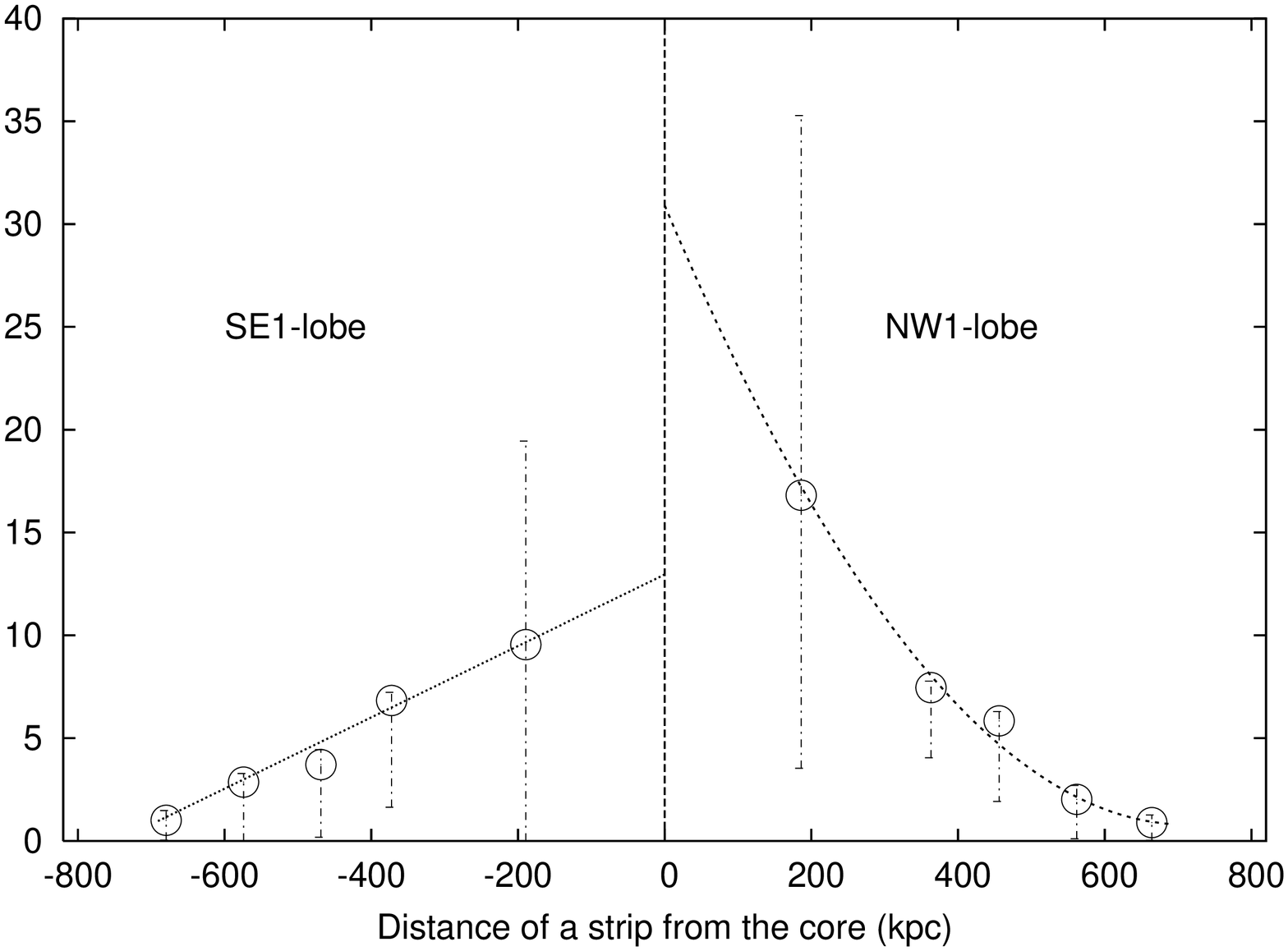,width=3.0in,angle=-90}
}
\caption[]{Radiative age of the relativistic particles in the outer lobes plotted against 
the distance from the radio core using the minimum energy magnetic fields. Top panel: Of the source 
J0041+3224. Bottom panel: Of the source J1835+6204. }
\label{age.dist.plot}
\end{figure}

To determine the spectral age of the particles in each strip, we
estimated the magnetic field strength of each strip. The values of the
minimum energy density and the corresponding magnetic field, B$_{\rm
  min}$, were calculated using the formalism described in Appendix of
Konar et al. (2008). We then used the values of the minimum energy
magnetic field and the break frequency (from
Table~\ref{age.strips_j0041} and \ref{age.strips_j1835}) to calculate
the spectral ages of the strips via Equation~\ref{eqn_specage} .
While performing spectral ageing analysis we have assumed that 1) the
energy losses of the relativistic cocoon plasma are only due to
synchrotron and IC processes, 2) the blob of plasma in each strip was
injected over a short period compared to the age of the radio galaxy,
so that the blob of plasma can be assumed to have been injected in a
single shot, 3) there is no mixing of plasma between any two adjacent
strips, 4) the magnetic field is in equipartition (the result of
Croston et al., 2005 is in favour of this assumption) and does not
change appreciably over the lifetime of the radio source and 5) the
magnetic field is completely tangled.
The spectral ages of the strips as a function of distance are plotted
in Figure~\ref{age.dist.plot}. As expected, the synchrotron ages for
both the outer lobes increase with distance from the edges
(warm-spots) of the lobes.  We have fitted a polynomial to every
age-distance plot to extrapolate the curve to the position of the
core.  The value of the polynomial at the position of the core gives
the expected spectral age of the outer lobes.  Since in both of our
DDRGs there is diffuse relativistic plasma from the outer lobes all
the way back to the core, it makes sense to determine the spectral age
by this extrapolation method. We could not determine the age of the
outer lobe plasma near the core region due to (i) the presence of
inner lobes and (ii) the non detection of diffuse plasma in our higher
frequency images. While interpreting these numbers, caveats related to
the evolution of the local magnetic field in the lobes need to be
borne in mind (e.g. Rudnick, Katz-Stone \& Anderson 1994; Jones, Ryu
\& Engel 1999; Blundell \& Rawlings 2000).  While Kaiser (2000)
suggested that spectral and dynamical ages are comparable if bulk
backflow and both radiative and adiabatic losses are taken into
account in a self-consistent manner, Blundell \& Rawlings (2000) argue
that this may be so only in young sources with ages much less than 10
Myr. In a study of the FRII type giant radio galaxy, J1343+3758,
Jamrozy et al. (2005) find the dynamical age to be approximately 4
times the maximum synchrotron age of the emitting particles.

\section{Discussion}
We find that the spectral ages of the two different outer lobes of the
DDRGs are different. This does not necessarily mean that the two jets
started at different points in time. This result may be due to
additional loss processes other than synchrotron and IC radiative
losses, (e.g., adiabatic expansion loss) and any kind of
re-acceleration of the particles that may be at work with differing
efficiency in the two lobes. The differing spectral ages then provide
information about the asymmetry of the physical conditions (internal
and external) of the lobes on opposite sides. A source can have
asymmetric jet powers, an asymmetric environment or both, in addition
to radiative losses and acceleration of the particles in the lobes.

In the western outer lobe of J0041+3224, we find that the variation of
spectral age with distance from the core is not smooth.  As we go from
the western hotspot towards the core we see that the spectral age
becomes almost constant from a distance of 400 kpc all the way to 175
kpc from the core.  This may be due to particle re-acceleration (of
unknown origin). The outer lobes of J0041+3224 are highly asymmetric
  which may be due to genuine asymmetry in the density of the
  ambient medium combined with asymmetric jet power and/or any other
  asymmetric parameters. Since the outer lobes are not active, the
  lobe heads are likely to advance with a speed slower than when they
  were active.  Since the radio spectra of the inner doubles are a
very close approximation to a single power law, we cannot determine
spectral ages for them.  However, we have determined upper limits to
the spectral ages of the inner doubles with the assumption that the
break frequencies are greater than the highest observed frequencies
(see Figure~\ref{spect_int.out.inn_j0041.n.j1835}).  We also
determined a lower limit to their age by assuming the jet-head
advancement speed to be 0.5 times the velocity of light.

As we have discussed, we find different radiative ages for the two
lobes of both sources. The actual ages of the lobes are very unlikely
to be different as it is not physically plausible for the two jets of
an AGN to start at different times. Therefore, our interpretation of
the apparent age difference is that processes other than radiative
losses, e.g., re-acceleration and adiabatic expansion, are at
work. Their relative impact on the energy loss in different lobes
might be different, giving rise to different radiative ages. We can
therefore obtain a more meaningful spectral age for the outer doubles
by taking an average of the spectral ages of the two lobes, provided
they are not different by huge factor (e.g.,$\ge$5 times). 
For J0041+3224, the spectral ages obtained by extrapolating
the spectral age vs. distance plot are $\sim$24 and 28 Myr for the NW1
and SE1 lobes respectively, with an average of 26 Myr. If we assume
that the re-acceleration of particles took place in the inner parts of
the NW1 lobe, then the extrapolated age of this lobe would be $\sim$84
Myr. In such a situation, we must invoke re-acceleration in the entire
SE1 lobe as it has a much lower spectral age than the NW1 lobe. It is
difficult to infer the true evolutionary history of this source.  For
J1835+6204, the extrapolated ages are $\sim$31 and 13 Myr for the NW1
and SE1 lobes, respectively, with an average of 22 Myr.  The average
spectral age of the two warm-spots ($t_{ws}$) is $\sim$4.44  Myr for
J0041+3224 and $\sim$1 Myr for J1835+6204.  The spectra of the inner
doubles show no curvature, and therefore we could not constrain the
break frequency or determine a spectral age.  However, the highest
frequencies in the power-law spectra of both the DDRGs are 8.46 GHz,
which can be assumed to be a lower limit on the break frequency. This
leads to upper limits on the spectral ages of the inner doubles of
$\sim$5.30 and 3.94 Myr for J0041+3224 and J1835+6204,
respectively.

It is known from various observations that the inner hotspots travel
with a velocity $\sim0.1c-0.5c$ with respect to the host galaxy (Konar
et al., 2006; Schoenmakers et al., 2000b; Safouris et al., 2008). So
we have estimated the lower limits on the ages (kinematic ages) of the inner
doubles for a  hotspot velocity of $\sim0.5c$, which are
$\sim$0.34 and 1.34 Myr for J0041+3224 and J1835+6204,
respectively. Of course, if one interprets the inner lobes to the bow 
shock driven by the jets (Brocksopp et al., 2007; 2011; Safouris et al., 2008) 
then the jet head velocity is supposed to be
close to the jet bulk velocity which will  be higher than 0.5$c$. In
that situation the lower limit of the ages of the inner doubles will 
be even lower. The spectral age of the outer hotspot is the time
elapsed since the last jet material reached the hotspot in the
previous episode of JFA; similarly the age of the inner double is the
time since the current episode of JFA started. So the durations of the
active phases ($t_{activ}$) and quiescent phase ($t_{quies}$) are given by
\begin{equation}
 t_{activ} = t_{outd} - ( t_{jet}+t_{ws} ),
\label{eqn_active.time}
\end{equation}
and 
\begin{equation}
 t_{quies} = (t_{ws} + t_{jet}) - t_{innd}
\label{eqn_quiescent.time}
\end{equation}
respectively; where, $t_{jet}$ is the time which any blob of jet material 
took to travel from the core to the warm-spot in the previous episode, 
$t_{outd}$ is the age of the outer double, and $t_{innd}$ is the 
age of the inner double. If we assume that the bulk velocity of the 
jet fluid is the velocity of light, then the values of $t_{jet}$ 
(when averaged over the two outer lobes) are 1.58 and 2.25 Myr for 
J0041+3224 and J1835+6204 respectively. So, the duration of active 
phase in the previous episode of JFA of J0041+3224 is $\sim$20 Myr 
(using Equation~\ref{eqn_active.time}). 
Since the estimated age of the  inner double of J0041+3224 is given by 
$0.34~\lapp t_{innd} \lapp~5.30$ Myr and $t_{jet}+t_{ws} = 6.02$ Myr, 
the duration of the quiescent phase
for this source is given by $0.72~\lapp t_{quies}\lapp~5.68$ Myr
(using Equation~\ref{eqn_quiescent.time}). So, quiescent phase of
J0041+3224 lies between 3.6 to 28.4 per cent of the active phase 
of the previous episode of JFA.  For J1835+6204 the situation is different.  
The break frequency is higher than the conventional fiducial value of the 
highest frequency ($\sim100$ GHz) up to which the spectrum exists (see
Table~\ref{age.strips_j1835}). Therefore, it is unlikely that there is
any real curvature in the spectrum. However, {\tt SYNAGE} tries its
best to fit with a curvature. There is no appreciable curvature in the
spectra of the outer hotspots (NW1-strip 01 and SE1-strip 01) of
J1835+6204 (see Figure~\ref{strip.fit_j1835}). Though we have formally
obtained an age of $\sim$1 Myr, ageing has not really started at the
outer hotspots, as the last ejected jet material is still feeding
them, which is particularly clear in the NW1 hotspot. Therefore, we 
will assume the ages of the outer hotspots to be zero in the sense 
that they are still fed, 
\begin{table}
\scriptsize{
\begin{center}
\caption{Results of JP model calculations for J0041+3224 with $\alpha_{\rm inj}=0.756^{+0.167}_{-0.122}$. 
Column~1: identification of the strip; column 2: the projected distance of the strip-centre from the radio core; 
column~3: the break frequency in GHz; column~4: the reduced $\chi^{2}$ value of the fit; column~5: the estimated 
magnetic field in nT; column~6: the resulting synchrotron age of the particles in the given strip.}
\label{age.strips_j0041}
\begin{tabular}{llllll}
\hline
Strip & Dist. & $\rm \nu_{br}$           & $\chi^{2}_{red}$  & $\rm B_{min}$  & $\rm \tau_{rad}$              \\
      & kpc   &    GHz                   &                   &       nT       & Myr                           \\
\hline

&&{\bf NW1-lobe} \\
NW1-01 &573.3&$ 3.68^{+1.62}_{-0.46}$     & 4.94              &1.86$\pm$0.18   & $3.68^{+1.62}_{-0.46}$        \\
NW1-02 &516.0&$ 5.76^{+3.27}_{-0.15}$     & 7.42              &1.73$\pm$0.17   & $5.76^{+3.27}_{-0.15}$      \\
NW1-03 &458.6&$11.11^{+1.91}_{-0.17}$     &29.95              &1.31$\pm$0.13   & $11.11^{+1.91}_{-0.17}$      \\
NW1-04 &401.3&$15.42^{+8.60}_{-0.57}$     &17.55              &1.12$\pm$0.11   & $15.42^{+8.60}_{-0.57}$     \\
NW1-05 &344.0&$14.50^{+6.49}_{-0.34}$     &30.75              &1.14$\pm$0.11   & $14.50^{+6.49}_{-0.34}$     \\
NW1-06 &286.6&$14.11^{+1.28}_{-0.42}$     & 1.70              &1.24$\pm$0.12   & $14.11^{+1.28}_{-0.42}$      \\
NW1-07 &229.3&$14.61^{+1.90}_{-0.31}$     &14.06              &1.16$\pm$0.11   & $14.61^{+1.90}_{-0.31}$      \\
NW1-08 &177.7&$15.72^{+1.45}_{-1.98}$     & 9.10              &1.07$\pm$0.11   & $15.72^{+1.45}_{-1.98}$      \\

&&{\bf SE1-lobe} \\
SE1-01 &378.4& $5.20^{+0.87}_{-0.34}$     &2.26               &2.01$\pm$0.20   & $5.20^{+0.87}_{-0.36} $      \\
SE1-02 &321.0& $6.81^{+0.61}_{-0.15}$     &8.01               &1.78$\pm$0.18   & $6.80^{+0.62}_{-0.14}$       \\
SE1-03 &263.7& $9.86^{+1.07}_{-0.17}$     &6.68               &1.42$\pm$0.14   & $9.86^{+1.07}_{-0.17}$       \\
SE1-04 &206.4& $12.23^{+0.47}_{-3.00}$    &1.31               &1.42$\pm$0.14   & $12.23^{+0.47}_{-3.00}$       \\
       &     &                            &                   &                &                               \\
\hline
\end{tabular}
\end{center}
}
\end{table}
\begin{table}
\scriptsize{
\begin{center}
\caption{Results of JP model calculations for J1835+6204 with $\alpha_{\rm inj}=0.818^{+0.070}_{-0.064}$. Column~1: identification
of the strip; column 2: the projected distance of the strip-centre from the radio core; column~3: the break 
frequency in GHz; column~4: the reduced $\chi^{2}$ value of the fit; column~5: the estimated magnetic field in nT;
column~6: the resulting synchrotron age of the particles in the given strip.} 
\label{age.strips_j1835}
\begin{tabular}{llllll}
\hline
Strip & Dist. & $\rm \nu_{br}$         & $\chi^{2}_{red}$ & $\rm B_{min}       $& $\rm \tau_{rad}$       \\
      & kpc   &      GHz               &                  & nT                  & Myr                    \\
\hline
&&{\bf NW1-lobe} \\
NW1-01 &664.2 & $243.5^{+1.34\times 10^5}_{-121.5}$     & 1.70         & 1.85$\pm$0.18       & $ 0.90^{+0.37}_{-0.86}$    \\
NW1-02 &561.7 & $123.5^{+3.94\time10^4}_{-54.7}$        & 2.39         & 1.21$\pm$0.12       & $ 2.02^{+0.69}_{-1.91}$   \\
NW1-03 &456.2 & $22.7^{186.3}_{-3.1}$                   & 2.90         & 0.94$\pm$0.09       & $ 5.85^{+0.44}_{-3.92}$   \\
NW1-04 &363.1 & $12.6^{+30.1}_{-1.0}  $                 & 4.02         & 1.00$\pm$0.10       & $ 7.64^{+0.14}_{-3.59}$  \\
NW1-05 &186.2 & $2.2^{+47.4}_{-1.7}  $                  & 0.38         & 1.08$\pm$0.11       & $16.82^{+18.46}_{-13.28}$   \\
       &      &                                         &              &                     &                           \\
&&{\bf SE1-lobe} \\
SE1-01 &679.7 & $421.0^{+1.96\times10^5}_{-225.0}$       &  1.84       &  1.31$\pm$0.13      & $1.01^{+0.47}_{-0.96}$   \\
SE1-02 &574.2 & $80.0^{+1.68\times10^{11}}_{-19.0}$      &  4.80       &  1.04$\pm$0.10      & $2.87^{+0.42}_{-2.87}$   \\
SE1-03 &468.6 & $82.6^{+3.14\times10^4}_{-24.6}$         & 14.43       &  0.67$\pm$0.07      & $3.72^{+0.72}_{-3.53}$   \\
SE1-04 &372.4 & $19.0^{+308}_{-2.0}$                     &  7.06       &  0.85$\pm$0.08      & $6.83^{+0.39}_{-5.18}$   \\
SE1-05 &189.3 & $9.58^{+1.81\times10^{11}}_{-7.27}$      &  0.99       &  0.86$\pm$0.08      & $9.54^{+9.90}_{-2.59}$   \\
       &      &                                          &             &                     &                           \\
\hline
\end{tabular}
\end{center}
}
\end{table}
and as a result the high energy particles are 
replenished. In high resolution  X band image (Figure~\ref{image.j1835_X-band}), 
we can easily see that there is a hotspot like feature in NW1 lobe. However
we don't see such prominant hotspot like feature in the SE1 lobe. Hotspot 
formation is a complicated phenomenon. It certainly seems possible that 
a diffuse hotspot can be formed without the jet having been turned off 
completely, and that is evident in the radio image of high resolution 
hotspots in some of the FR\,II sources published by Leahy et al., (1997). 
Moreover, if there is a weak compact hotspot like the one in north western 
lobe of 3C105 (see Leahy et al., 1997), we might not be able to see 
because of the sensitivity of our high resolution X band image in 
Figure~\ref{image.j1835_X-band}. So, it is difficult to conclude whether 
the SE1 lobe is still fed by the jets or not. Here we argue from both 
points of view.

First we assume that NW1 and SE1 lobes are still fed by the jets of previous episode. 
The last jet material has been ejected from the nucleus, but the jet material is still 
travelling down the jets. However, we do not know how long ago the last jet material was 
ejected from the nucleus. We assume that $t^{'}_{jet}$, which is 
$\lapp t_{jet}$ ($=2.25$ Myr), is the time since the last jet material was ejected. 
So to calculate $t_{activ}$ and $t_{quies}$, $t_{jet}$ and $t_{ws}$ should be 
replaced by $t^{'}_{jet}$ and 0 in Equations~\ref{eqn_active.time} and 
\ref{eqn_quiescent.time} respectively. 
However, $t^{'}_{jet}$ must be greater than the lower limit of $t_{innd}$ as the last 
jet material of previous cycle had to be ejected before the current episode started. 
So $t^{'}_{jet} \gapp~1.34$. Again, if the last ejected jet material has already reached 
the hotspots, then $t^{'}_{jet}= t_{jet}=2.25$ Myr. So, we ultimately get 
$1.34~\lapp t^{'}_{jet} \lapp~2.25$ Myr. The upper limit of the age of the inner double 
can be at most $t_{jet}$, not greater than that. So, $t_{innd} \le t_{jet}$. Therefore,
we get $1.34~\lapp t_{innd} \lapp~2.25$ Myr. Considering that $t_{quies}$ must be a 
positive number, and the limits of $t_{innd}$ and $t^{'}_{jet}$,  
we get $19.75~\lapp t_{activ}\lapp~20.66$ Myr and $0~\lapp t_{quies} \lapp~0.91$ Myr. 
The limits of $t_{activ}$ are so close to each other that the average of those limits 
is a good approximation of the $t_{activ}$ which is $\sim$20.20 Myr.
So, in this range of quiescent phase, the last ejected jet material can travel from the 
nucleus up to 279 kpc, and not more than that. So the quiescent phase of J1835+6204 
has lasted less than $\sim4.5$ per cent of the active phase of the previous episode of 
JFA.

If we assume that the lack of a hotspot in SE1 lobe is real, then it may suggest that 
jet activity has ceased there and we expect that the hotspots of the inner double are 
the current jet termination points. In such a case, the presence of a compact hotspot 
in the NW1 lobe is the feature that needs to be explained. Since hotspots of the inner 
double are the current jet termination points, then we would expect the hotspot in
the NW1 lobe to disappear at some point. Assuming also that the jet turns off on both 
sides at the same time, the only way we can see a hotspot in the NW1 lobe and not in the 
SE1 lobe is if the NW1 lobe is on the far side of the source, so that we are seeing it 
at an earlier time. For the sake of a simple calculation we assume that the source is 
symmetrical. Let the true lobe length (from the core to the hotspot) be $L$ and the 
angle to the line of sight be $\theta$. Then $L_{obs} = 2Lsin\theta$ and the distance 
along the line of sight between the hotspots is given by $D_{LTT} = 2Lcos\theta$, where 
$L_{obs}$ is the observed lobe length and $D_{LTT}$ is basically the Light Travel Time (LTT) 
multiplied by speed of light ($c$). So $D_{LTT}$ is $L_{obs}cot\theta$. If we require 
$\theta > 45$ degrees (unification arguments) then the maximum value of 
$D_{LTT} = L_{obs}\sim$1.4 Mpc. So the maximum time difference between the hotspots is 
4.6 Myr. Suppose the jet was disconnected from the lobes at a time $t^{'}_{jet}$ in 
the past and the effects travel down the old jet channel at a speed $\beta_{jet}c$. If 
we see no hotspot in the SE lobe, then we know that $\beta_{jet}ct^{'}_{jet} > L$. But 
if we see a hotspot in the NW1 lobe, which we see at an earlier time, then we know 
$\beta_{jet}c(t^{'}_{jet} - D_{LTT}/c) < L$, 
or $\beta_{jet}ct^{'}_{jet} < (L + \beta_{jet} D_{LTT})$. For simplicity if we assume 
$\theta = 45$ degrees and $\beta_{jet} = 1$, then we have $ct^{'}_{jet}\gapp 1$ Mpc
and $ct^{'}_{jet}\lapp (1 + 1.4)$ Mpc. So, essentially we obtain
$3.3~\lapp t^{'}_{jet} \lapp~7.9$ Myr. So the jet switch-off time has to be more than 
3.3 Myr ago and less than 7.9 Myr ago (measured with respect to the observations of the SE1 hotspot). 
We have already found out that the jet travel time $t_{jet}$ of the last ejected material 
for J1835+6204 is 2.25 Myr which is less than $t^{'}_{jet}$. Therefore, we can calculate 
the hotspot fading away time ($t_{hs,fade}$) from this analysis. The hotspot fading time 
can be given by
\begin{equation}
t_{hs,fade}= t_{ws}= t^{'}_{jet} - t_{jet}.
\label{eqn_fadeing.time}
\end{equation}   
Considering the limits of $t^{'}_{jet}$ for the source J1835+6204, we obtain 
$1.05~\lapp t_{hs,fade}\lapp~5.65$ Myr. Even if the hotspot fade away time is the lower limit
of this range, i.e., 1.05 Myr which is enough for the SE1 hotspot to get diffuse, as it takes 
only sound crossing time which is $\sim10^5$ yr (Kaiser et al., 2000). From 
Equation~\ref{eqn_fadeing.time}, the age of the warm spot ($t_{ws}$) in SE1 lobe can be given 
by $1.05~\lapp t_{ws} \lapp~5.65$ Myr. Using the magnetic field value of $1.31\pm0.13$ nT of 
the SE1-strip01 from Table~\ref{age.strips_j1835} into Equation~\ref{eqn_specage}, we obtain 
$13.5 ~\lapp \nu_{br}\lapp~ 390$ GHz. So, our estimation of warmspot age in the SE1 lobe of 
J1834+6204 in our interpretation of absence of the SE1 hotspot is fully consistent with the 
fact that we don't see any curvature in the spectrum of SE1-strip01 (see Figure~\ref{strip.fit_j1835}) 
in our observable range of frequencies. This particular set of numbers doesn't seem to make 
it so implausible that we are seeing a genuine one-sided hotspot, though these are the most 
favourable assumptions. As $\theta$ approaches 90 degrees, we have to be observing the source 
at more and more special times to see a hotspot in one lobe and not in the other, since our 
constraints are basically $1 < \frac{ct_{jet}}{L} < 1 + 2 cos\theta$. So, for any source with 
roughly symmetrical lobes there will be some time window where we expect to see a hotspot in 
one lobe but not in the other. Now again in Equations~\ref{eqn_active.time} and \ref{eqn_quiescent.time}, 
we replace $t_{jet}$ and $t_{ws}$ by $t^{'}_{jet}$ and 0 (as the spectrum is straight) respectively. 
Keeping in mind the limits of $t^{'}_{jet}$ and $t_{innd}$, we obtain from Equations~\ref{eqn_active.time} 
and \ref{eqn_quiescent.time} that $14.1 ~\lapp t_{activ}\lapp~18.7$ Myr and $1.05 ~\lapp t_{quies}\lapp~6.56$ Myr. 
So, if this interpretation is correct, then the quiescent phase of J1835+6204 can be given by
$5.6~\lapp ~t_{quies}~\lapp ~46.5$ percent of the active phase of the previous cycle.  

\section{Concluding remarks}
We have presented the results of multifrequency radio observations of two DDRGs, J0041+3224 and J1835+6204, 
using both the GMRT and the VLA. We have carried out detailed spectral ageing analysis of these two sources, and 
have come to the following conclusions.   
\begin{enumerate}
\item From our observations, it is quite clear that the inner doubles are propagating through the cocoon material 
      deposited in the previous episode of JFA. This may be the case for most of the DDRGs, even though the old 
      cocoon material is not always visible in images of limited sensitivity.

\item Our spectral ageing analysis reveals the youth of the inner
  doubles and the old age of the outer doubles, consistent with the
  model in which the central engine is capable of restarting a
  previously quenched jet of the central engines, and with all
  previous work.

\item The images show that in both cases the pairs of doubles are not
  significantly misaligned. This suggests that there has not been any
  realignment of the galaxy gas disk, accretion disk and/or jet axis
  of the central engine.

\item The radio core of J1835+6204 shows no evidence of variability,
  whereas Konar et al. (2006) and Jamrozy et al. (2007) have reported
  high variability in the cores of the restarting radio galaxies
  J1453+3308 and 4C29.30 respectively.  The results we present here
  suggest that strong core variability may or may not be observed in
  restarting radio galaxies.

\item The synchrotron age of the emitting particles in both the lobes
  of the outer doubles increases with distance from the edges of the
  lobes, as expected from the backflow model of FRII radio
  galaxies ({\bf This behaviour is also seen in lobed FR\,I sources. See  
  Parma et al., 1999 and Laing et al., 2011. }). 
  There appears to be either re-acceleration of particles in
  the inner part of the NW1 lobe of J0041+3224, or adiabatic loss in
  the middle of this lobe. The reasons for the re-acceleration or
  adiabatic loss are not known. Further investigation is required to
  understand this phenomenon.

\item The duration of the quiescent phase of J0041+3224 is greater than 
  0.72 Myr but less than 5.68 Myr, which is between $\sim4-28$ per cent 
  of the duration of the active phase of the previous episode of JFA, provided 
  the age of the outer double is 26 Myr. If the age of the outer double is 84 Myr,
  then the duration of quiescent phase is $\sim0.9-7$ per cent of the duration of 
  the active phase of the previous episode of JFA. 
  The duration of the quiescent phase of J1835+6204 is less than 
  $\sim$1 Myr, which is $\sim4.5$ per cent of the active phase of the previous 
  episode of JFA. It is therefore clear that the quiescent phase can 
  be as small as a few per cent of the active phase duration.   

\item Our age estimates of J1835+6204 are fully consistent with the outer lobes 
  being still fed by the last ejected jet material, though the alternative
  explanation of one sided visible hotspot because of the LTT effect 
  is also physically plausible.  
\end{enumerate}

\section*{Acknowledgments} 
We thank the GMRT and the VLA staffs for their help with the observations. The Giant Metrewave 
Radio Telescope is a national facility operated by the National Centre for Radio Astrophysics 
of the Tata Institute of Fundamental Research. The National Radio Astronomy Observatory  is a 
facility of the National Science Foundation operated under co-operative agreement by Associated 
Universities Inc. This research has made use of the NASA/IPAC extragalactic database which 
is operated by the Jet Propulsion Laboratory, Caltech, under contract with the National Aeronautics
and Space Administration. CK and MJ acknowledge the access to the {\tt SYNAGE} software 
provided by M. Murgia (Istituto di Radioastronomia, Bologna, Italy. MJ acknowledges the Polish 
MNiSW funds for scientific research in the years 2009-2012 under agreement no. 3812/B/H03/2009/36. 
JHC acknowledges support from the South-East Physics Network (SEPNet). CK acknowledges the grant 
(No. NSC99-2112-M-001-012-MY3) from the National Science Council, Taiwan.

{}
\end{document}